\documentclass[5p, twocolumn]{elsarticle}[draft]

\usepackage{color}
\usepackage{lineno}
\usepackage[pagebackref=true,breaklinks=true,letterpaper=true,colorlinks,bookmarks=true]{hyperref}
\usepackage[ruled,vlined,linesnumbered]{algorithm2e}
\usepackage{multirow}
\usepackage{booktabs}       % professional-quality tables
\usepackage{subfigure}
\usepackage{xspace}
\usepackage{float}
\usepackage{blindtext}
\usepackage{url}
\usepackage{amsmath}
\usepackage{amssymb}
\usepackage{amsfonts}
\usepackage{multicol}
\usepackage{graphicx}
\usepackage{array}
\usepackage{makecell}
\usepackage{float}
\usepackage{xurl}
\usepackage{bookmark}
\usepackage{easyReview}

\makeatletter
\DeclareRobustCommand\onedot{\futurelet\@let@token\@onedot}
\def\@onedot{\ifx\@let@token.\else.\null\fi\xspace}

\def\eg{\emph{e.g}\onedot} 
\def\ie{\emph{i.e}\onedot}

\def\etal{\emph{et al}\onedot}
\makeatother

\SetKwInput{KwInput}{Input}                % Set the Input
\SetKwInput{KwOutput}{Output}

\journal{Journal of Neurocomputing}

\begin{document}

\begin{frontmatter}

    \title{Deep Plug-and-Play Prior for Hyperspectral Image Restoration}
    \author[a]{Zeqiang~Lai}
    %\ead{laizeqiang@bit.edu.cn}
    \author[a]{Kaixuan~Wei}
    %\ead{kaixuan_wei@bit.edu.cn}
    \author[a]{Ying~Fu\corref{cor1}}
    \ead{fuying@bit.edu.cn}

    \cortext[cor1]{Corresponding author}

    \address[a]{School of Computer Science and Technology, Beijing Institute of Technology, Beijing, 100081, China.}

    \begin{abstract}
        Deep-learning-based hyperspectral image (HSI) restoration methods have gained great popularity for their remarkable performance but often demand expensive network retraining whenever the specifics of task changes.
        In this paper, we propose to restore HSIs in a unified approach with an effective plug-and-play method, which can jointly retain the flexibility of optimization-based methods and utilize the powerful representation capability of deep neural networks. Specifically, we first develop a new deep HSI denoiser leveraging gated recurrent convolution units, short- and long-term skip connections, and an augmented noise level map to better exploit the abundant spatio-spectral information within HSIs. It, therefore, leads to the state-of-the-art performance on HSI denoising under both Gaussian and complex noise settings.
        Then, the proposed denoiser is inserted into the plug-and-play framework as a powerful implicit HSI prior to tackle various HSI restoration tasks.
        Through extensive experiments on HSI super-resolution, compressed sensing, and inpainting, we demonstrate that our approach often achieves superior performance, which is competitive with or even better than the state-of-the-art on each task, via a single model without any task-specific training.
    \end{abstract}


\end{frontmatter}

\section{Introduction}
Hyperspectral image (HSI) consists of numerous discrete bands from a wide range of continuous spectrums for each spatial location and provides rich spectral information beyond the visible, which makes it exceedingly useful in the applications of face recognition \cite{1251148, zhao2014sparse}, remote sensing \cite{blackburn2007hyperspectral,thenkabail2016hyperspectral,zhang2010object}, classification \cite{azar2020hyperspectral,zhou2019hyperspectral,cao2019hyperspectral}, and more. However, due to the cost and complexity of hyperspectral imaging, HSIs are often limited to low spatial resolution and suffered from different types of degradations, such as noise, missing pixels, and undersampling. As a result, restoring clean, high-resolution, and complete hyperspectral data becomes a crucial initial step for the success of subsequent HSI applications.

In general, HSI restoration can be considered as an inverse imaging problem, \ie reconstructing the original HSI $x$ from the degraded observation $y$ that is obtained via a forward process $y=D(x)+e$, where $D$ is the degradation operation, and $e$ is assumed to be additive Gaussian white noise (AWGN). Many tasks, such as HSI denoising, super-resolution, inpainting, and compressed sensing, can fit into this framework by specifying $D$ as identity, downsampling, masking, and sensing operation, respectively. However, the problems are that most of these forward processes are ill-posed, which means the corresponding inverse problems are difficult or even impossible to be solved without extra prior knowledge about HSIs.

To address the aforementioned problems, classical HSI restoration approaches minimize a cost function that consists of a data-fidelity term, which measures how well the reconstructed HSIs match the observations, and a regularization term, which reflects certain prior knowledge with respect to unknown HSIs. Following this methodology, numerous handcrafted priors are developed, \eg total variation \cite{yuan2012hyperspectral,wang2017hyperspectral}, low-rank tensor modeling \cite{chang2020weighted, chang2017hyper}, and adaptive spatial-spectral dictionary learning \cite{fu2017adaptive}. These methods are generally flexible in solving multiple restoration tasks with tiny adjustments and able to achieve reasonable performance under certain assumptions about input data. However, they also suffer from two serious drawbacks. First, their performance is inherently restricted by the matching degree of handcrafted priors and intrinsic properties of HSIs. Second, most of these methods are computationally expensive as they require the iterative minimization of the cost function for each input HSI.

With the emergence of deep learning, many works \cite{wei20203,yuan2018hyperspectral} have been developed to circumvent the design of handcrafted priors by taking advantage of the powerful representation capability of deep neural networks. Such methods directly learn the mapping from degraded observations to reconstructed HSIs from a large number of training pairs and implicitly embed the learned prior knowledge into the parameters of the neural network. In this way, not only do they achieve the state-of-the-art performance in many ill-posed HSI restoration tasks \cite{hu2020hyperspectral,mei20173dfcnn,wong2020hsi}, but they also reduce tremendous inference time in comparison with classical methods. Despite the strong superiority, deep-learning-based methods literally lose the flexibility to tackle different restoration tasks in one model. One has to design and train separate models for different tasks or even identical tasks with slightly different settings, which is cumbersome.

As an alternative, the commonly used plug-and-play (PnP) framework \cite{venkatakrishnan2013plug,zhang2020plug} in Gray/RGB image restoration seems to be a promising choice for its capability of leveraging the flexibility of optimization-based methods and the data-driven prior by replacing the traditional handcrafted regularizer with a deep-learning one. However, unlike the conventional Gray/RGB images that mainly contain spatial information, HSIs provide richer spectral information that should be considered, which makes the direct use of Gray/RGB denoisers, such as FFDNet \cite{zhang2018ffdnet}, unsuitable as they generally ignore the important spatial-spectral correlation and the global correlation along the spectrum in HSIs. Besides, as a general solution to various image restoration tasks, the PnP method often requires different denoising strengths to achieve desirable performance under different settings, so it is better for PnP denoisers to be capable of handling a wide range of continuous noise levels, which is also missing in most existing HSI denoising algorithms.

In this paper, we propose to leverage the plug-and-play framework with a novel deep HSI denoiser to solve multiple HSI restoration problems in a unified approach. In detail, we first use the alternating direction method of multipliers (ADMM) \cite{boyd2011distributed} to decouple the traditional optimization objective into two independent subproblems where one subproblem can be solved in closed-form and another subproblem related to image prior can be implicitly solved by an off-the-shelf denoiser. Then, a deep HSI denoiser, \ie a gated recurrent convolutional neural network (GRCNN), is introduced to circumvent the shortcomings of existing commonly used Gray/RGB plug-and-play denoisers. Specifically, a gated recurrent convolution block is adopted as the basic building block to exploit the rich spectrum correlation in HSIs, and its effectiveness is further enhanced by a residual encoder-decoder architecture. Meanwhile, an additional noise level map is employed to help to train a robust denoiser that is able to handle a wide range of continuous noise levels. Over extensive experiments, it is shown that our deep HSI denoiser not only outperforms the existing HSI denoising algorithms, but also supports the proposed plug-and-play method achieving the superior generalizability and performance against the learning-based and optimization-based methods, respectively, in the tasks of HSI super-resolution, compressed sensing, and inpainting.

In summary, our main contributions are that:
\begin{itemize}
    \item A deep Plug-and-Play ADMM approach is presented to solve the HSI restoration tasks in a unified approach, in which the prior modeling is implicitly handled by a deep HSI denoiser.
    \item A new deep HSI denoiser is introduced to better exploit the intrinsic characteristics of HSIs via gated recurrent convolution units, short- and long-term skip connections, and an auxiliary noise level map, meanwhile supporting the proposed plug-and-play HSI restoration framework.
    \item Extensive experiments on four typical HSI restoration tasks, including denoising, super-resolution, compressed sensing, and inpainting, demonstrate that our plug-and-play approach is able to effectively and flexibly tackle a variety of HSI restoration problems without any task-specific training.
\end{itemize}

The rest of this paper is organized as follows. Section \ref{sec:related-work} reviews the related works of HSI restoration and plug-and-play. Section \ref{sec:method} presents the detailed illustration of the proposed plug-and-play HSI restoration method and the proposed deep denoiser. Section \ref{sec:experiment} provides the experimental results for four HSI restoration tasks and gives a series of discussions. Section \ref{sec:conclude} concludes the paper with a brief summary.

\section{Related Works}\label{sec:related-work}
In this section, we review the relevant studies on HSI restoration and plug-and-play methods.
\subsection{HSI Restoration}

Traditional optimization-based HSI restoration methods usually solve an inverse imaging problem with extra regularizations by exploiting the underlying characteristics of HSIs. By considering the spectral correlation, many works, such as total variational methods \cite{yuan2012hyperspectral,wang2017hyperspectral}, wavelet methods \cite{othman2006noise}, and low-rank models \cite{zhao2020fast,sun2017hyperspectral,wang2017hyperspectral,wei2019low}, have been developed. The non-local self-similarity is another important property of HSIs and was exploited in works like block-matching and 4-D filtering (BM4D) \cite{maggioni2012nonlocal} and the tensor dictionary learning \cite{peng2014decomposable}. With the development of deep learning, many learning-based methods were proposed, \eg Wei \etal \cite{wei20203} presented a novel network QRNN3D for HSI denoising, Mei \etal \cite{mei20173dfcnn} introduced a 3D fully convolution neural network for HSI super-resolution. Although these methods achieved remarkable results in their respective fields, most of them are specifically developed for a single task.

Recently, a number of works \cite{chang2020weighted,fu2017adaptive,he2019non,pengE3DTV} have been developed to solve multiple HSI restoration problems in a unified approach. For example, Chang \etal \cite{chang2020weighted} utilized the low-rank tensor prior to model the spatial non-local self-similarity and spectral correction simultaneously and applied their method to HSI denoising, deblurring, destriping, and super-resolution. By considering the sparsity on the subspace bases on gradient maps, Peng \etal \cite{pengE3DTV} proposed an enhanced 3D total variation prior for both HSI denoising and compressed sensing. In the work of Fu \etal \cite{fu2017adaptive}, a novel adaptive spatial-spectral dictionary learning method was introduced for HSI denoising and super-resolution. Despite the superior performance and flexibility, they are time-consuming and heavily rely on how well the handcrafted prior matches with the intrinsic properties of HSIs.

\subsection{Plug-and-Play Methods}

Since introduced in \cite{chan2016plug,danielyan2010image, zoran2011learning}, the idea of plug-and-play has received great attention for its flexibility and effectiveness to solve a wide range of inverse imaging problems. In the area of traditional Gray/RGB image restoration, there have been many attempts \cite{danielyan2010image,sun2020block,zhang2020plug,zhang2017learning,zhang2019deep} for injecting denoising priors into the plug-and-play framework. In \cite{danielyan2010image}, Danielyan \etal combined the augmented Lagrangian technique with a BM3D prior \cite{dabov2007image}.
By leveraging DnCNN \cite{zhang2017beyond} denoiser as the implicit regularizer, Sun \etal \cite{sun2020block} developed a block coordinate regularization-by denoising (RED) algorithm. Inspired by the design of FFDNet \cite{zhang2018ffdnet}, Zhang \etal \cite{zhang2020plug} proposed a DRUNet that adopts a noise level map as additional input for plug-and-play denoising. The recent work in \cite{zhang2019deep}, from a new perspective, proposed to use a deep super-resolver instead of denoiser as the implicit prior for plug-and-play single image super-resolution. Although all the denoisers \cite{zhang2017beyond, zhang2017learning,zhang2018ffdnet} for plug-and-play Gray/RGB image restoration can be directly extended to the HSI cases, none of them specifically explore the extensive domain knowledge of HSIs.

Some recent works \cite{liu2021hyperspectral, ma2020hyperspectral} also attempt to inject plug-and-play deep denoising prior for HSI restoration. In \cite{liu2021hyperspectral}, Liu \etal employed BM3D \cite{dabov2007image} as an extra plug-and-play regularization for its fibered rank constrained tensor restoration framework. In \cite{ma2020hyperspectral}, FFDNet \cite{zhang2018ffdnet} was used for local regularity, where the global structures are constrained by Kronecker-basis-representation-based tensor low-rankness. However, the plug-and-play priors in these works are still Gray/RGB denoisers that cannot regularize the global spatial-spectral correlation, which limits the performance of these methods for extremely ill-posed HSI restoration tasks, such as inpainting and compressed sensing. On the contrary, by considering both local and global information simultaneously, the single 3D denoiser we used can not only accelerate the iterative process but also benefit the ill-posed HSI restoration by rich prior knowledge encoded in the network. We refer interested readers to \cite{wei2020tfpnp} for a thorough review of the plug-and-play algorithms.

\section{Deep Plug-and-Play HSI Restoration} \label{sec:method}

We consider the degradation model for HSI restoration as
\begin{equation}
    y=Dx+e,
\end{equation}
where $x \in \mathbb{R}^{MNB}$ is the latent clean image, $y\in \mathbb{R}^{M_hN_hB}$ is the degraded observation, $D$ is the noise-irrelevant degradation matrix, $e$ is assumed to be additive white Gaussian noise and $M,N,B$ denote the number of columns, rows, and bands of HSIs, respectively.

In the following, we first introduce the plug-and-play ADMM algorithm in the proposed deep plug-and-play framework. Then, a deep HSI denoiser acting as the implicit image prior is presented.

\subsection{Plug-and-Play ADMM}\label{sec:admm}

The main idea of plug-and-play ADMM is to separate the prior term and data-fidelity term of the traditional optimization objective into two subproblems, where the subproblem related to prior can be treated as a denoising problem that can be solved by an off-the-shelf denoiser, \eg a deep HSI denoiser.
According to this, we can not only retain the flexibility to solve multiple HSI restoration problems in a unified model, but also leverage the powerful prior-modeling ability provided by deep neural networks.

In detail, the adopted PnP-ADMM solves the following regularized objective:
\begin{equation}
    (\widehat{{x}}, \widehat{{v}})=\underset{{x}}{\operatorname{argmin}}  \enspace \frac{1}{2} || {D}{x}- {y} ||^{2}+\lambda g({v}), \quad \text{\emph{s.t. }} {x}={v},
    \label{eq-admm-decouple}
\end{equation}
where $g$ denotes the regularization and $\lambda$ is a non-negative weighting term. By considering its augmented Lagrangian function:
\begin{equation}
    \begin{aligned}
        \mathcal{L}({x}, {v}, {u})= & \frac{1}{2} || {D}{x} - {y} ||^{2} +\lambda g({v})  +{u}^{T}({x}-{v}) \\&+\frac{\rho}{2}\|{x}-{v}\|^{2}.
    \end{aligned}
    \label{eq-admm-unconstrain}
\end{equation}
Equation \eqref{eq-admm-decouple} can be solved by minimizing $\mathcal{L}$, which is achieved by alternately solving the following three subproblems:
\begin{subequations}
    \begin{align}
        {x}^{(k+1)}       & = \underset{{x}}{\operatorname{argmin}} \enspace \frac{1}{2} || {D}{x} - {y} ||^{2}+\frac{\rho}{2}\left\|{x}-\tilde{{x}}^{(k)}\right\|^{2} \label{eq-admm-x}, \\
        {v}^{(k+1)}       & = \underset{{v} \in \mathbb{R}^{n}}{\operatorname{argmin}} \enspace g({v})+\frac{1}{2 \sigma^{2}}\left\|{v}-\tilde{{v}}^{(k)}\right\|^{2} \label{eq-admm-v},  \\
        \bar{{u}}^{(k+1)} & = \bar{{u}}^{(k)}+\left({x}^{(k+1)}-{v}^{(k+1)}\right),
    \end{align}
    \label{eq-admm}
\end{subequations}
where $\bar{{u}}^{(k)} \equiv (1/\rho) {u}^{(k)}$ is the scaled dual variable of ${v}$ and $\rho$ is a penalty parameter, $\tilde{{x}}^{(k)} \equiv {v}^{(k)} - \tilde{{u}}^{(k)}$, $\tilde{{v}}^{(k)} \equiv {x}^{(k+1)} + \bar{{u}}^{(k)}$ and $\sigma \equiv \sqrt{\lambda / \rho}$.

\paragraph{Update $v$} Equation \eqref{eq-admm-v} can be considered as solving a Gaussian HSI denosing problem, where $\sigma$ is the noise level, ${v}$ is the "clean" image, and $\tilde{{v}}^{k}$ is the corresponding "noisy" image. To explain, we could first consider the MAP estimate of "clean" image $v$:
\begin{equation}
    v = \underset{{x}}{\operatorname{argmin}} \enspace \{-\ln p(\tilde{{v}}^{k}|v) - \ln p(v)\}.
    \label{eq:map}
\end{equation}
If we assume the noise is AWGN with the variance of $\sigma^2$, then we have
\begin{equation}
    -\ln p(\tilde{{v}}^{k}|v) = \frac{1}{2 \sigma^{2}}\left\|{v}-\tilde{{v}}^{(k)}\right\|^{2} + \ln(\sigma \sqrt{2\pi}).
    \label{eq:map2}
\end{equation}
By substituting Equation \eqref{eq:map2} into Equation \eqref{eq:map} and considering $-\ln p(v)$ as $g(v)$, Equation \eqref{eq-admm-v} is technically equivalent with Equation \eqref{eq:map}, which proves the fact that Equation \eqref{eq-admm-v} can be considered as a Gaussian denoising problem.

As a result, we could practically solve Equation \eqref{eq-admm-v} with any off-the-shelf HSI denoising algorithm $\mathcal{D}_{\sigma}$, \eg a deep HSI denoiser.
\begin{equation}
    {v}^{(k+1)}=\mathcal{D}_{\sigma}\left(\widetilde{{v}}^{(k)}\right).
\end{equation}

\paragraph{Update $x$} As for the subproblem in Equation \eqref{eq-admm-x}, the general solution could be obtained by solving a least square problem in  the closed-form, \ie,
\begin{equation}
    x ^{(k+1)} = \left( D^TD +\rho I \right)^{-1}\left( D^Ty +\rho\tilde{x}^{(k)} \right).
    \label{eq:admm-x-general}
\end{equation}

For the super-resolution task, $D=SH$ where $H$ is a circulant matrix denoting the blur operation. $S$ is a binary matrix denoting the $k$ times  downsampling.
The fast solution of Equation \eqref{eq:admm-x-general} is given by \cite{chan2016plug}, which is expressed as:
\begin{equation}
    {x}=\rho^{-1} {b}-\rho^{-1} {G}^{T}\left(\mathcal{F}^{-1}\left\{\frac{\mathcal{F}({G} {b})}{\left|\mathcal{F}\left(\widetilde{h}_{0}\right)\right|^{2}+\rho}\right\}\right),
\end{equation}
where $\mathcal{F}$ and $\mathcal{F}^{-1}$ denote Fast Fourier Transform (FFT) and inverse FFT, ${G}={S H}$, ${b}={G}^{T} {y}+\rho \widetilde{{x}}$, and $\tilde{h}_0$ is the $0^{th}$ polyphase component of the filter $HH^T$.

For the compressed sensing task, $D = \boldsymbol{\Phi} \in R^{MN\times MNB}$ is the sensing matrix. The fast solution of Equation \eqref{eq-admm-x} can be found in \cite{liu2018rank}, which is shown below:
\begin{equation}
    \boldsymbol{x}^{(k+1)} = \boldsymbol{x}^{(k)} +\boldsymbol{\Phi}^{\top} \left[\frac{y-\boldsymbol{\Phi}\boldsymbol{\tilde{x}}^{(k)}}{\operatorname{diag}\left\{\rho+\psi_{1}, \ldots, \rho+\psi_{n}\right\}} \right],
\end{equation}
where $\Phi \Phi^{\top} \stackrel{\text { def }}{=} \operatorname{diag}\left\{\psi_{1}, \ldots, \psi_{n}\right\}$.

For the inpainting task, $D=S$ is a diagonal masking matrix.
The solution is exactly the same as the general solution in Equation \eqref{eq:admm-x-general}, but we could implement matrix inverse as element-wise division.
\begin{equation}
    x ^{(k+1)} = \left( S^TS +\rho I \right)^{-1}\left( S^Ty +\rho\tilde{x}^{(k)} \right).
\end{equation}
Specifically, since $S$ is diagonal, $S^T = S$ and $S^TS$ are diagonal, which makes $(S^TS +\rho I)$ diagonal as well. Hence, the matrix inversion of $(S^TS +\rho I)^{-1}$ can be implemented as element-wise division. Furthermore, $S^T = S$, $S^Ty=Sy$ can be viewed as a masking process and can be efficiently implemented as element-wise multiplication.

Following the similar derivations described above, our PnP framework can also be used for other tasks, \eg, HSI deblurring \cite{liao2013hyper,wang2020hyper}, HSI and MSI fusion \cite{li2018fusing, dian2018deep}. In short, solving HSI deblurring is identical with HSI super-resolution with the scale factor set to 1, while HSI-MSI fusion needs some modifications of the regularized objective shown in Equation \eqref{eq-admm-decouple}.

\paragraph{Parameter Setting} There are two parameters that must be set for each iteration in the PnP-ADMM algorithm, \ie, the penalty parameter $\rho$ and the noise level of denoiser $\sigma$.  We adopt a similar strategy presented in \cite{zhang2020plug} to set these parameters. In detail, $\sigma$ is uniformly sampled from a large noise level $\sigma_1$ to a small one $\sigma_2$ in log space. $\rho$ is determined by the relation $\sigma=\sqrt{\lambda / \rho}$ where $\lambda$ is empirically set to $1.5$. It is noted that the performance may be further improved by tuning the parameters for each image with recent method \cite{wei2020tuning} based on reinforcement learning.

The summary of the proposed deep plug-and-play HSI restoration algorithm can be found in Algorithm \ref{alg:pnp}.

\begin{algorithm}
    \small
    \SetAlgoLined
    \KwInput{Degraded HSI ${y}$, degradation operator $D$, parameters $\rho$ and $\sigma$ for each iteration, number of iterations $K$ and deep denoiser $\mathcal{D}$}
    \KwOutput{Restored HSI ${x}$}
    Perform task-specific initialization of $x$\;
    Initialize $v = x$, $u = 0$\;
    \For{$k=1,2,...,K$}{
    ${x}^{(k+1)} = \underset{{x}}{\operatorname{argmin}} \frac{1}{2} || {D}\tilde{{x}}^{(k)} - {y} ||^{2}+\frac{\rho}{2}\left\|{x}-\tilde{{x}}^{(k)}\right\|^{2}$\;
    ${v}^{(k+1)}=\mathcal{D}_{\sigma}\left(\widetilde{{v}}^{(k)}\right)$\;
    $\bar{{u}}^{(k+1)} = \bar{{u}}^{(k)}+\left({x}^{(k+1)}-{v}^{(k+1)}\right)$\;
    }
    \caption{Deep Plug-and-Play HSI Restoration with Alternating Direction Method of Multipliers.}
    \label{alg:pnp}
\end{algorithm}

\subsection{Deep HSI Denoiser}

\begin{figure*}[t]
    \begin{center}
        \includegraphics[width=0.97\linewidth]{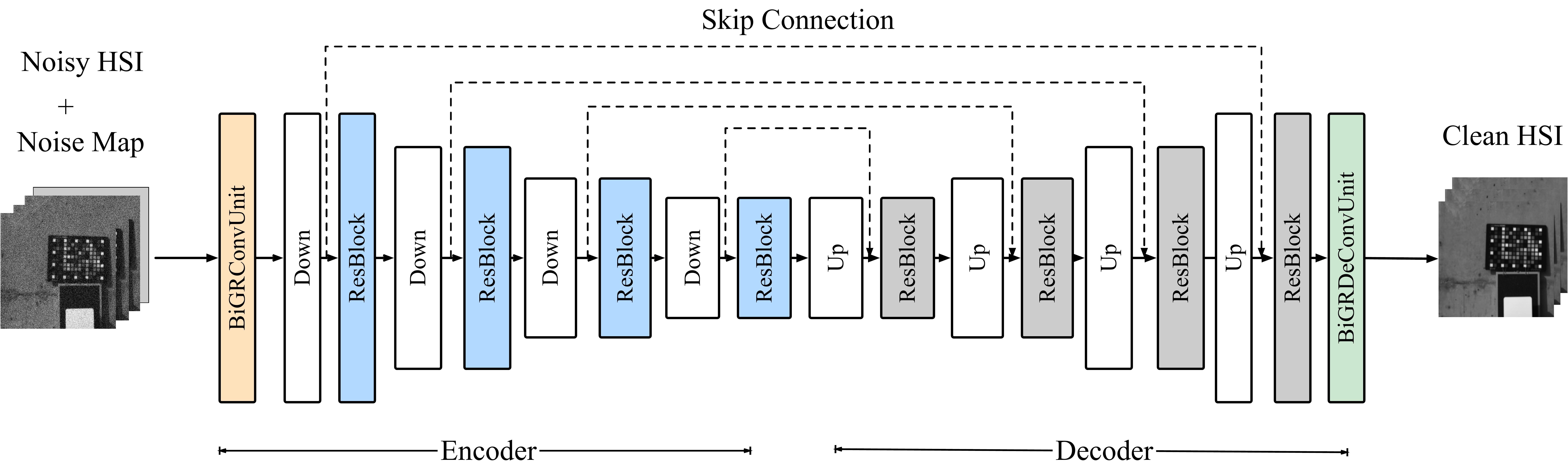}
    \end{center}
    \caption{The overall architecture of our deep plug-and-play denoising prior, where the first and the last blocks are two bidirectional GRConv units, the Down/Up block refers to downsample/upsample GRConv unit, and the ResBlock is composed of several GRConv units with residual connections that keep the spatial dimension but increase the number of features.}
    \label{fig:dgrunet}
\end{figure*}

In contrast to traditional Gray/RGB image denoising task, HSI denoising for plug-and-play has its unique requirements and challenges. First, HSI is made up of a massive number of bands in a wide range of spectrum that correlate with each other and should be considered in the process of denoising. As such, direct use of Gray/RGB denoisers, such as FFDNet \cite{zhang2018ffdnet}, DnCNN \cite{zhang2017beyond}, and IRCNN \cite{zhang2017learning}, would not achieve decent performance due to the lack of the exploration of spatial-spectral correlation in HSIs. Second, most existing denoisers are only able to handle a or a set of fixed noise levels, which is not suitable for the plug-and-play framework. In the following, we present our \emph{Gated Recurrent Convolutional Neural Network (GRCNN)} to address the aforementioned issues.

\paragraph{Network Architecture}
Inspired by the commonly used UNet \cite{ronneberger2015u} and the recently proposed QRNN3D \cite{wei20203} architectures, we design our gated recurrent convolutional neural network as a deep encoder-decoder to exploit the complex information underlying HSIs. As shown in Figure \ref{fig:dgrunet}, the encoder consists of repeated application of a downsample gated recurrent convolution (GRConv) unit to decrease the spatial size and a GRConv residual block to increase the number of features. The decoder is symmetrically set up, and every step in the expansive path includes an upsample GRConv unit and a residual block to reconstruct the clean HSIs. At the first and final layers, we use two bidirectional GRConv units to handle the mapping between input/output HSIs and extracted features. Meanwhile, long-term skip connections in the form of concatenation are also used to inject shallow features from the encoder into the decoder to guide the reconstruction of clear images.

\paragraph{Gated Recurrent Convolution Unit}

\begin{figure}[h!]

    \begin{center}
        \includegraphics[width=0.95\linewidth]{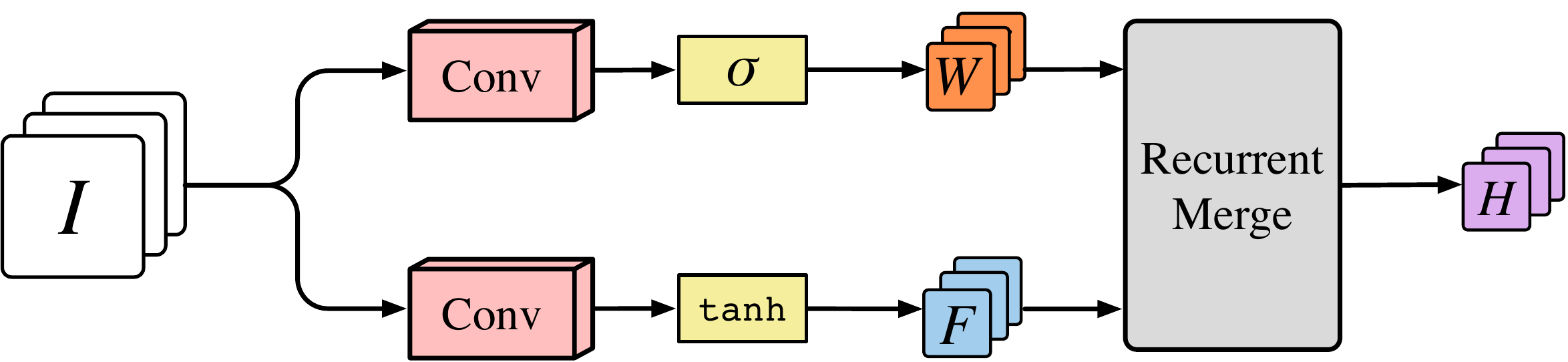}
    \end{center}

    \caption{The overall structure of the gated recurrent convolution unit. $I$, $W$, $F$, $H$ denote the input feature maps, weight maps, candidate feature maps, and merged feature maps, respectively.}
    \label{fig:grconv}
\end{figure}

Intuitively, if we assume the noise is randomly and independently added to each individual pixel at each band, it is highly possible that one noisy pixel in one band remains relatively clean in another band. Such intuition suggests that if we can somehow know the quality of pixels in different bands at the same spatial location, we might be able to utilize the clean pixels to guide the denoising of the noisy ones via the global spectral correlation. Based on the above discussion, we introduce a gated recurrent convolution (GRConv) unit to replace the traditional convolutional layer as the basic building block in our network, and its overall structure is shown in Figure \ref{fig:grconv}.

In detail, the proposed GRConv first separately performs two 3D convolution/deconvolution on input features $I$ to obtain a set of pixel-wise weight maps $W$ and feature maps $F$ band by band, as described in Equation \eqref{eq:grc-conv} where $h_w, h_f$ are two 3D filters, $\otimes$ represents the convolution, and $\sigma$ is the sigmoid activation. The weight map ranges from 0 to 1 and acts as an indicator telling what percentage of information in the corresponding feature map we should keep.
\begin{equation}
    \begin{aligned}
        W & = \sigma \left(h_w \otimes I \right) \\
        F & = \tanh \left(h_f \otimes I \right)
    \end{aligned}
    \label{eq:grc-conv}
\end{equation}

Then, a weighted merging step is applied to fuse the candidate features among the bands recurrently as shown in Equation \eqref{eq:grc-rnn}:
\begin{equation}
    h_i = (1-w_i) \odot h_{i-1} + w_i \odot f_i, \,\,\, \forall  i\in [1,b],
    \label{eq:grc-rnn}
\end{equation}
where $\odot$ denotes the element-wise multiplication and $w_i$, $f_i$, $h_i$ denote the weight map, the candidate feature map and the fused feature map at the $i^{th}$ band, respectively.

It should be noted that Equation \eqref{eq:grc-rnn} only merges the features in one direction. In the bidirectional GRConv unit, we perform the same fusion with different weight maps in another direction and stack the fused feature maps in both directions to form the final feature maps. To reduce the computational complexity, all the GRConv units except the first and the last ones are single directional, and we alternatively change the merging directions to make sure each band can receive guidance from both directions.

\paragraph{Residual Block}  With the network goes deeper and deeper, it becomes increasingly difficult to stabilize the training and make the model converge to a desirable local minimum. Inspired by the ResNet \cite{he2016deep}, we build our basic network component as a residual block which consists of two $3\times3\times3$ gated recurrent convolution/deconvolution units and one projection shortcut implemented with a $1\times1\times1$ GRConv unit to keep the shallow information flow. The shortcuts together with the skip connections enable both the short- and long-term information interaction, thereby enlarging the representation capacity of the network.

\begin{figure}[h!]
    \begin{center}
        \includegraphics[width=0.6\linewidth]{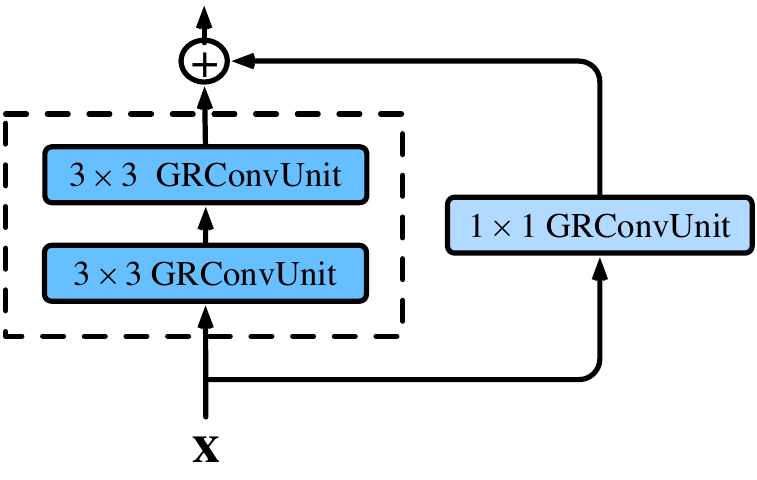}
    \end{center}
    \caption{The illustration of Resblock. It contains two $3\times3$ GRConv unit and one $1\times1$ GRConv unit to increase the number of feature maps, while keeping the spatial resolution unchanged.}
    \label{fig:residual-block}
\end{figure}

\paragraph{Noise Level Map} As indicated in Equation \eqref{eq-admm-v}, the denoiser for our PnP-ADMM algorithm should be a Gaussian denoiser and be able to handle a relatively wide range of continuous noise levels. Existing Gray/RGB/HSI denoisers are not ideal for this task as most of them are designed for only a small set of noise levels. Although we can directly train the network with a training dataset covering a range of continuous noise levels, such a strategy lacks  effective guidance, which might impose an extra burden on the network for inferencing the noise strength. With the aim of improving the generalizability to remove noise at continuous levels, we propose to train the denoiser with noise level map as an additional input, which acts as a teacher that directly tells the network the demanded denoising strength.

\section{Experiments}\label{sec:experiment}

In this section, we first evaluate the proposed deep HSI denoiser on Gaussian and complex HSI denoising (Section \ref{sec:denoise}). Then, without any additional training, the proposed denoiser is plugged into our plug-and-play framework to solve the other three classical HSI restoration tasks. The experimental results in super-resolution, compressed sensing, and inpainting are provided in Section \ref{sec:sr}, Section \ref{sec:cs} and Section \ref{sec:inpainting}, respectively.
The detailed ablation study of our approach is presented in Section \ref{sec:abl} in the final.

\subsection{Experimental Settings} \label{sec:dataset}

\paragraph{Dataset}
We train the proposed denoiser with the natural hyperspectral image dataset, \ie, ICVL \cite{arad2016sparse}, which consists of 201 images with 31 spectral bands in the size of $1392 \times 1300$. We randomly select 100 images for training, 5 images for validation, and 50 images for testing. Each image in the training set is cropped into multiple overlapped volumes of size $64\times64\times31$ to enlarge the dataset. Besides, data augmentation techniques such as rotation and scaling are also adopted, resulting in roughly 50k training samples in total. The main region of each test image with the size of $512\times 512\times 31$ is used for testing.

\paragraph{Implementation Details}
We implement the proposed denoising network in PyTorch \cite{paszke2019pytorch} . The Adam \cite{kingma2014adam} optimizer with default parameters is adopted to minimize the L2 loss between reconstructed HSI and ground truth. In order to obtain a robust denoiser for our plug-and-play framework, the network is first trained on a fixed noise level of 50 for 30 epochs and then fine-tuned on random noise levels ranging from 0 to 50 for another 20 epochs. This version of the trained denoiser is used for the experiments of HSI super-resolution, compressed sensing and inpainting. When compared with other gaussian denoising methods, we further fine-tune our model with noisy images randomly corrupted by four fixed noise levels, \ie 10, 30, 50, 70, to obtain the model for the evaluation on Gaussian denoising task. When compared with other complex denoising methods, our network is trained without the noise level map using the same strategy of Gaussian noise for the first 50 epochs and then fine-tune on complex noise for another 50 epochs. The initial learning rate is set to $10^{-3}$ and gradually decayed to stabilize the training. It takes roughly two days for a complete training with an NVIDIA RTX 3090 GPU.

Without any additional training, the pretrained denoiser can be plugged into our plug-and-play framework to solve other HSI restoration tasks, but the best hyperparameters vary for different tasks. In order to achieve the best performance, we empirically choose the best hyperparameters of our PnP-ADMM algorithm for HSI super-resolution, compressed sensing, and inpainting tasks. Besides, the inputs for the PnP-ADMM are initialized with bicubic interpolation and triangulation-based linear interpolation for HSI super-resolution and inpainting, respectively.

\paragraph{Quantitative Metrics} To evaluate the performance of the proposed method comprehensively, three quantitative quality indices are used, \ie, PSNR, SSIM \cite{wang2004image}, and SAM \cite{yuan2018hyperspectral}. PSNR and SSIM measure the spatial quality while SAM focuses on the spectral quality. The larger PSNR and SSIM values the better the restore images are, while the smaller values of SAM imply better performance. PSNR and SSIM are calculated as the average of the bandwise results for each HSI.

\begin{table*}[h!]
    \caption{Quantitative denoising results of different methods under several noise levels on ICVL dataset. [30, 70] suggests each image is corrupted by Gaussian noise with random $\sigma$ (ranged from 30 to 70).}
    \vspace{3mm}
    \small
    \begin{center}
        \begin{tabular}{@{}cccccccccc@{}}
            \toprule
            \multirow{2}{*}{Sigma}    & \multicolumn{1}{c}{\multirow{2}{*}{Metric}} & \multicolumn{8}{c}{Method}                                                                                                                                                                                                        \\ \cmidrule(l){3-10}
                                      & \multicolumn{1}{c}{}                        & Noisy                      & BM4D\cite{maggioni2012nonlocal} & KBR\cite{Qi2017Kronecker} & WLRTR\cite{chang2020weighted} & NGmeet\cite{he2019non} & HSID-CNN\cite{yuan2018hyperspectral} & QRNN3D\cite{wei20203} & Ours           \\ \midrule
            \multirow{3}{*}{30}       & PSNR                                        & 18.59                      & 38.45                           & 41.48                     & 42.62                         & 42.99                  & 38.70                                & 42.22                 & \textbf{43.06} \\
                                      & SSIM                                        & 0.110                      & 0.934                           & 0.984                     & 0.988                         & 0.989                  & 0.949                                & 0.988                 & \textbf{0.990} \\
                                      & SAM                                         & 0.807                      & 0.126                           & 0.088                     & 0.056                         & 0.050                  & 0.103                                & 0.062                 & \textbf{0.050} \\ \bottomrule
            \multirow{3}{*}{50}       & PSNR                                        & 14.15                      & 35.60                           & 39.16                     & 39.72                         & 40.26                  & 36.17                                & 40.15                 & \textbf{40.91} \\
                                      & SSIM                                        & 0.046                      & 0.889                           & 0.974                     & 0.978                         & 0.980                  & 0.919                                & 0.982                 & \textbf{0.984} \\
                                      & SAM                                         & 0.991                      & 0.169                           & 0.100                     & 0.073                         & 0.059                  & 0.134                                & 0.074                 & \textbf{0.059} \\ \bottomrule
            \multirow{3}{*}{70}       & PSNR                                        & 11.23                      & 33.70                           & 36.71                     & 37.52                         & 38.66                  & 34.31                                & 38.30                 & \textbf{38.82} \\
                                      & SSIM                                        & 0.025                      & 0.845                           & 0.961                     & 0.967                         & 0.974                  & 0.886                                & 0.974                 & \textbf{0.976} \\
                                      & SAM                                         & 1.105                      & 0.207                           & 0.113                     & 0.095                         & \textbf{0.067}         & 0.161                                & 0.094                 & 0.086          \\ \bottomrule
            \multirow{3}{*}{[30, 70]} & PSNR                                        & 17.34                      & 37.66                           & 40.68                     & 41.66                         & 42.23                  & 37.80                                & 41.37                 & \textbf{42.23} \\
                                      & SSIM                                        & 0.114                      & 0.914                           & 0.979                     & 0.983                         & 0.985                  & 0.935                                & 0.985                 & \textbf{0.987} \\
                                      & SAM                                         & 0.859                      & 0.143                           & 0.087                     & 0.064                         & \textbf{0.053}         & 0.116                                & 0.068                 & 0.056          \\ \bottomrule
        \end{tabular}
    \end{center}
    \label{tab:denoise-gaussian}
\end{table*}

\begin{figure*}[h!]
    \centering
    \subfigure[BM4D]{
        \includegraphics[width=0.85in]{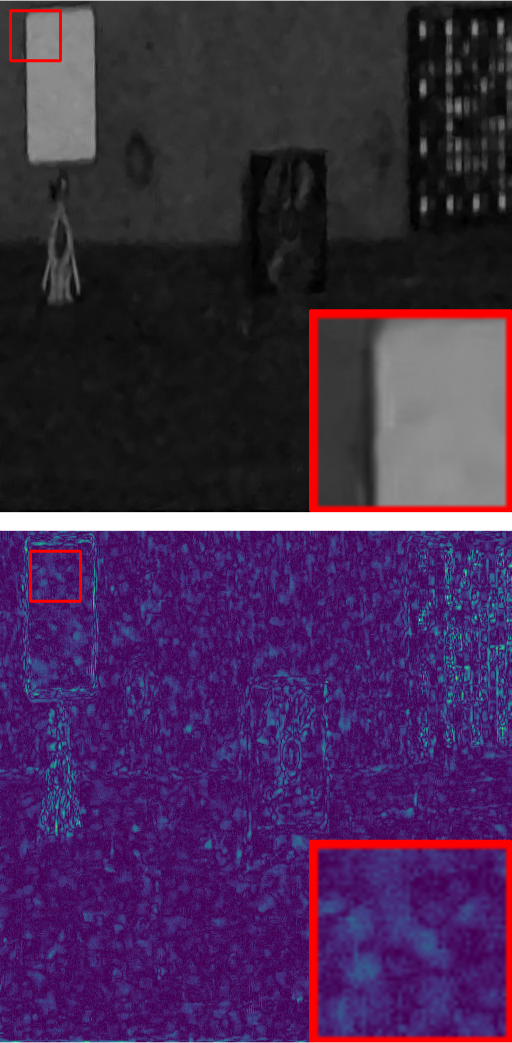}
    }
    \subfigure[KBR]{
        \includegraphics[width=0.85in]{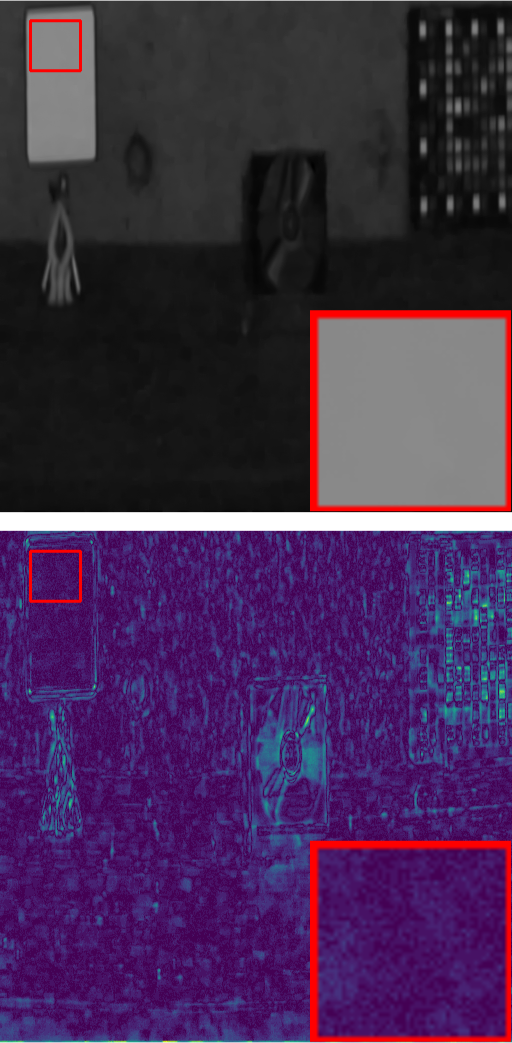}
    }
    \subfigure[WLRTR]{
        \includegraphics[width=0.85in]{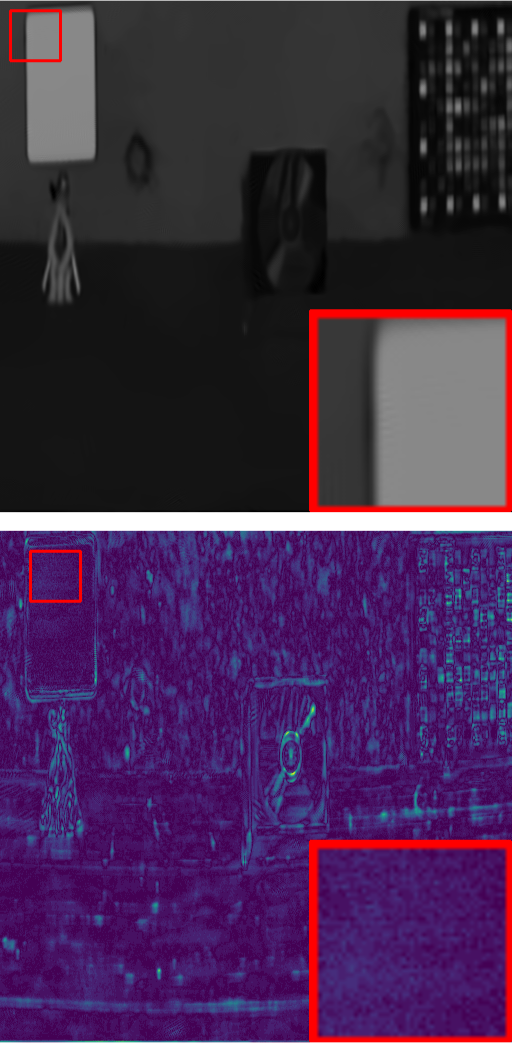}
    }
    \subfigure[NGmeet]{
        \includegraphics[width=0.85in]{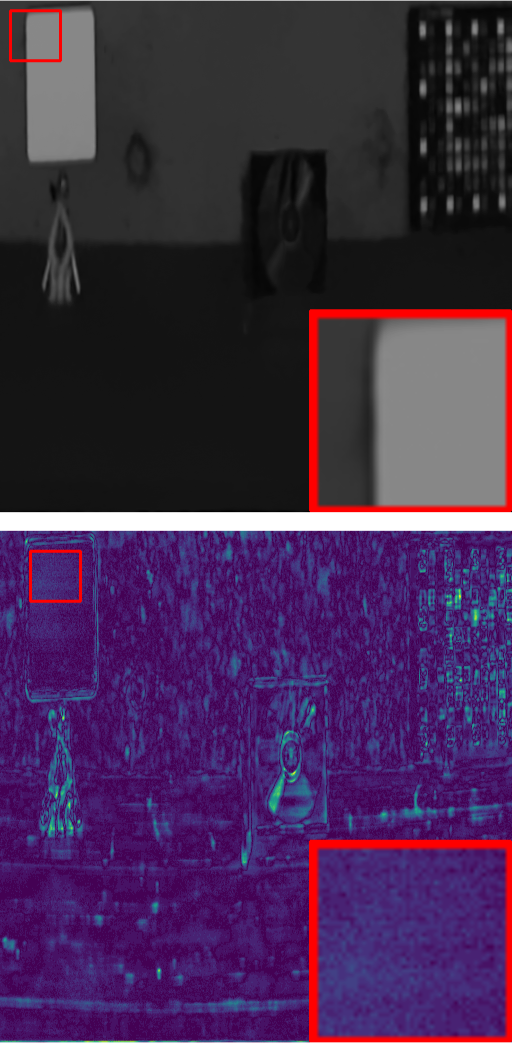}
    }
    \subfigure[HSID-CNN]{
        \includegraphics[width=0.85in]{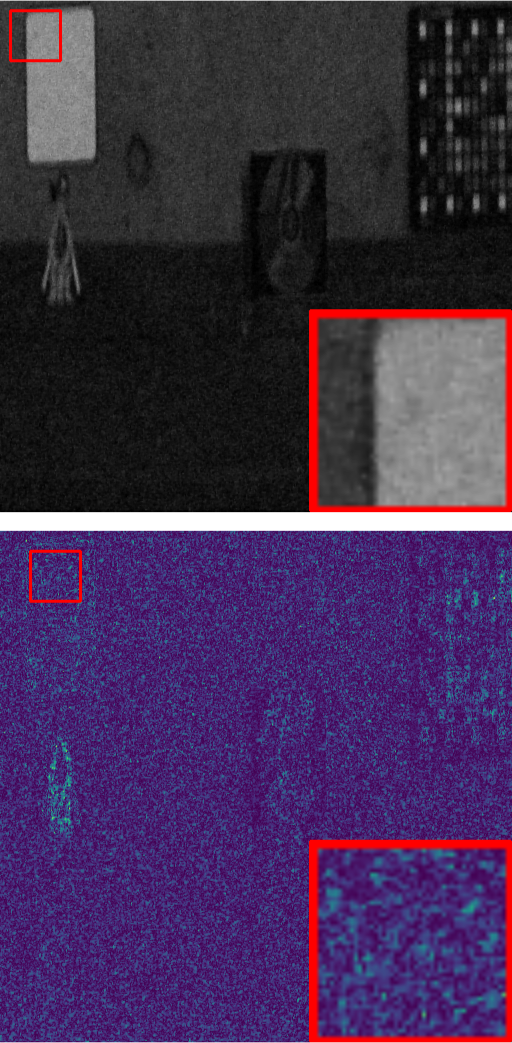}
    }
    \subfigure[QRNN3D]{
        \includegraphics[width=0.85in]{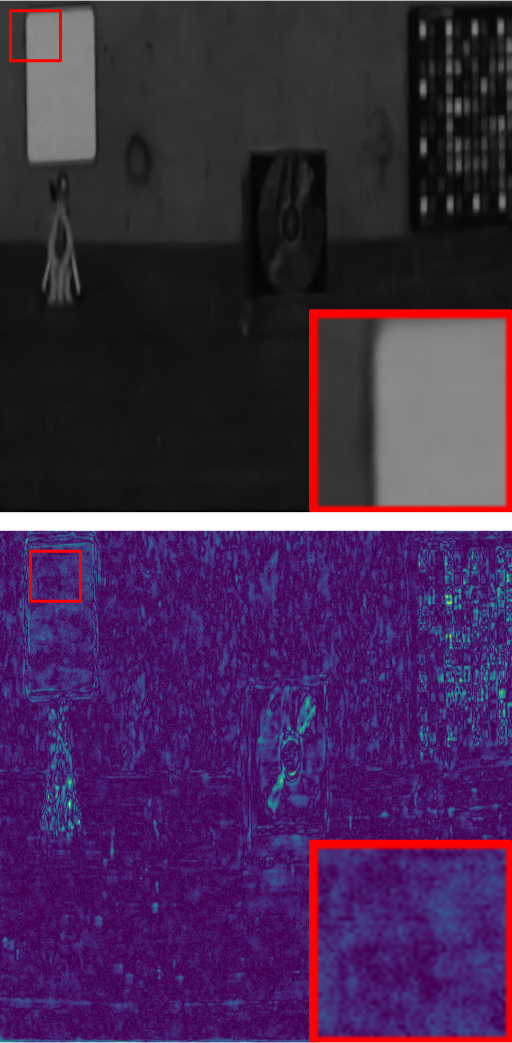}
    }
    \subfigure[Ours]{
        \includegraphics[width=0.85in]{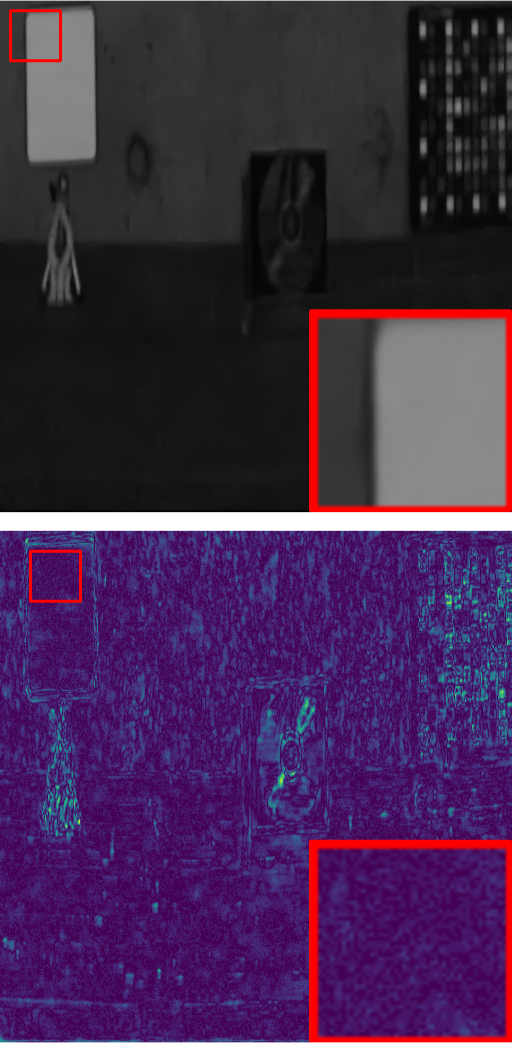}
    }
    \vspace{3mm}
    \caption{Simulated Gaussian noise removal results at the $20^{th}$ band of images under noise level $\sigma=50$ on ICVL dataset. The first row shows the denoising results. The second row includes the corresponding error maps. The error maps are the absolution errors between the ground truth and the recovered results. The brighter position indicates the higher error.}
    \label{fig:denoising-gaussian}
\end{figure*}

\begin{table*}[h!]
    \caption{Quantitative denoising results of different methods under three complex noise cases on ICVL dataset. \emph{non-iid} denotes Non i.i.d Gaussian noise. \emph{g+stripe} denotes Non i.i.d Gaussian noise and stripe noise. \emph{g+impulse} denotes Non i.i.d Gaussian noise and impulse noise. (Please refer to Section \ref{sec:exp-complex} for the detailed explanation.)}
    \vspace{3mm}
    \small
    \begin{center}
        \begin{tabular}{@{}cccccccccc@{}}
            \toprule
            \multirow{2}{*}{Sigma}     & \multicolumn{1}{c}{\multirow{2}{*}{Metric}} & \multicolumn{8}{c}{Method}                                                                                                                                                                                                                \\ \cmidrule(l){3-10}
                                       & \multicolumn{1}{c}{}                        & Noisy                      & LRMR\cite{zhang2013hyperspectral} & LRTV\cite{he2015total} & NMoG\cite{chen2017denoising} & TDTV\cite{wang2017hyperspectral} & HSID-CNN\cite{yuan2018hyperspectral} & QRNN3D\cite{wei20203} & Ours           \\ \midrule
            \multirow{3}{*}{non-iid}   & PSNR                                        & 18.25                      & 32.80                             & 33.62                  & 34.51                        & 38.14                            & 38.40                                & 42.79                 & \textbf{42.89} \\
                                       & SSIM                                        & 0.168                      & 0.719                             & 0.905                  & 0.812                        & 0.944                            & 0.947                                & 0.978                 & \textbf{0.992} \\
                                       & SAM                                         & 0.898                      & 0.185                             & 0.077                  & 0.187                        & 0.075                            & 0.095                                & 0.052                 & \textbf{0.047} \\ \bottomrule
            \multirow{3}{*}{g+stripe}  & PSNR                                        & 17.80                      & 32.62                             & 33.49                  & 33.87                        & 37.67                            & 37.77                                & 42.35                 & \textbf{42.39} \\
                                       & SSIM                                        & 0.159                      & 0.717                             & 0.905                  & 0.799                        & 0.940                            & 0.942                                & 0.976                 & \textbf{0.991} \\
                                       & SAM                                         & 0.910                      & 0.187                             & 0.078                  & 0.265                        & 0.081                            & 0.104                                & 0.055                 & \textbf{0.050} \\ \bottomrule
            \multirow{3}{*}{g+impulse} & PSNR                                        & 14.80                      & 29.70                             & 31.56                  & 28.60                        & 36.67                            & 35.00                                & 39.23                 & \textbf{40.70} \\
                                       & SSIM                                        & 0.114                      & 0.623                             & 0.871                  & 0.652                        & 0.935                            & 0.899                                & 0.945                 & \textbf{0.985} \\
                                       & SAM                                         & 0.926                      & 0.311                             & 0.242                  & 0.486                        & 0.094                            & 0.174                                & 0.109                 & \textbf{0.067} \\ \bottomrule
        \end{tabular}
    \end{center}
    \label{tab:denoise-commplex}
\end{table*}

\begin{figure*}[ht!]
    \centering
    \setlength{\tabcolsep}{0.09cm}
    \begin{tabular}{cccccccc}
         & LRMR                                                                                                    & LRTV & NMoG & TDTV & HSID-CNN & QRNN3D & Ours \\
        \rotatebox{90}{\quad non-iid}
         & \multicolumn{1}{m{0.12\linewidth}}{\includegraphics[width=0.85in]{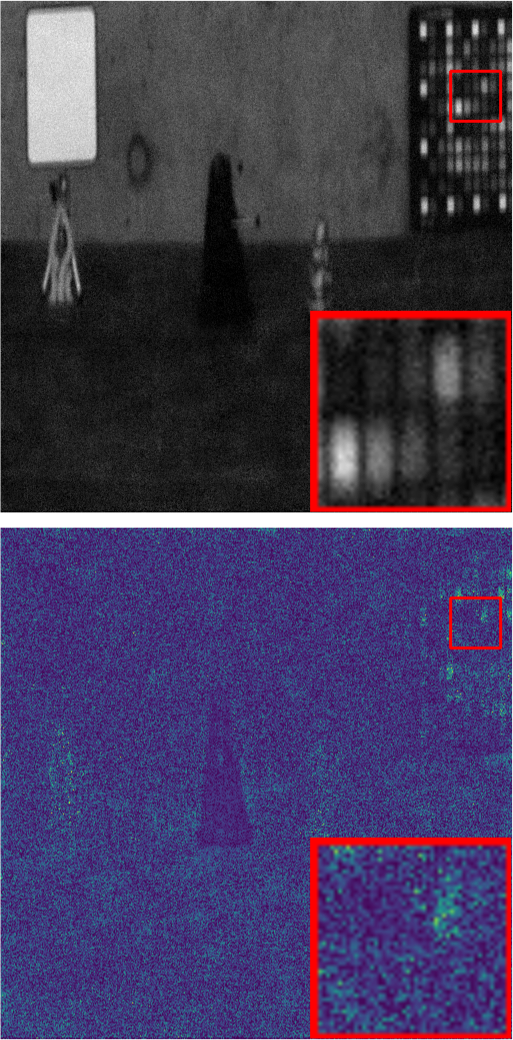}}
         & \multicolumn{1}{m{0.12\linewidth}}{\includegraphics[width=0.85in]{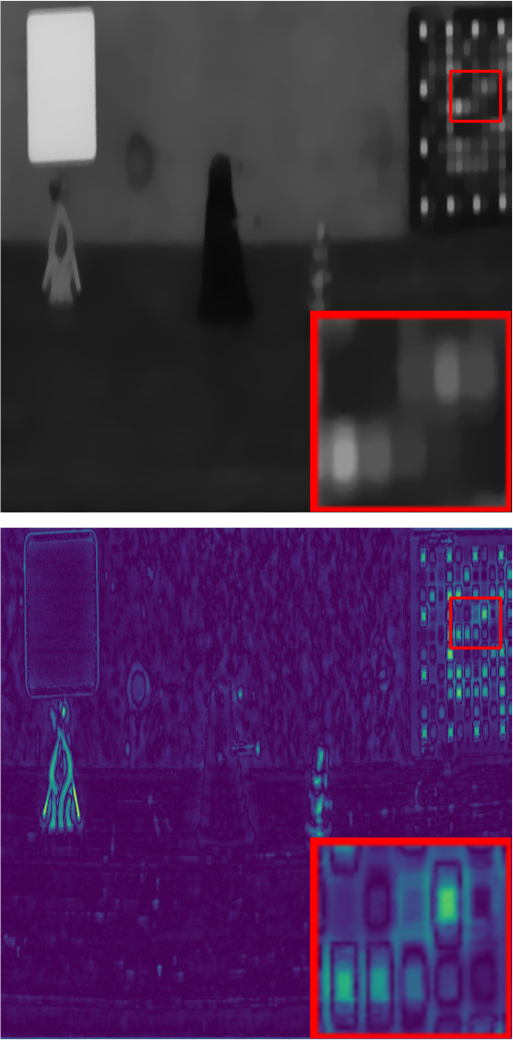}}
         & \multicolumn{1}{m{0.12\linewidth}}{\includegraphics[width=0.85in]{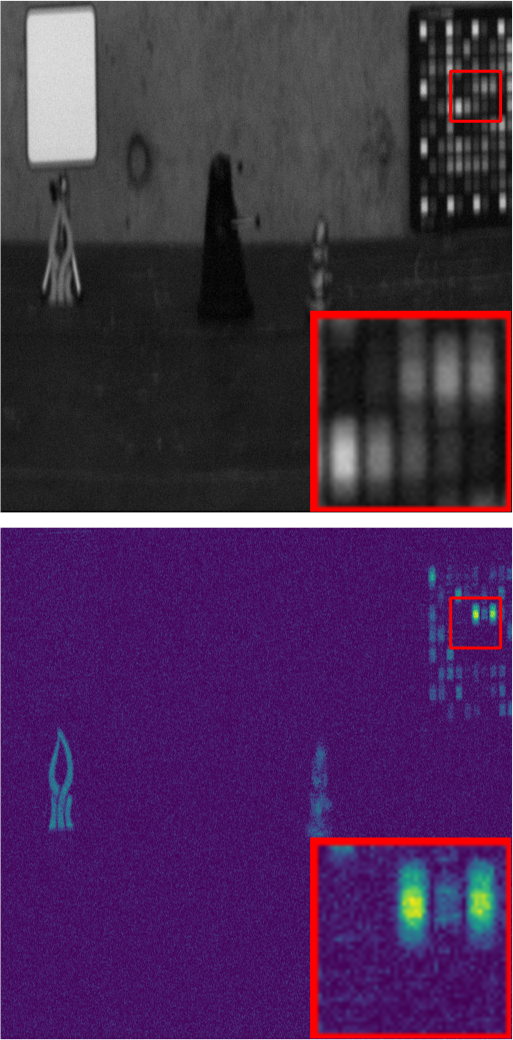}}
         & \multicolumn{1}{m{0.12\linewidth}}{\includegraphics[width=0.85in]{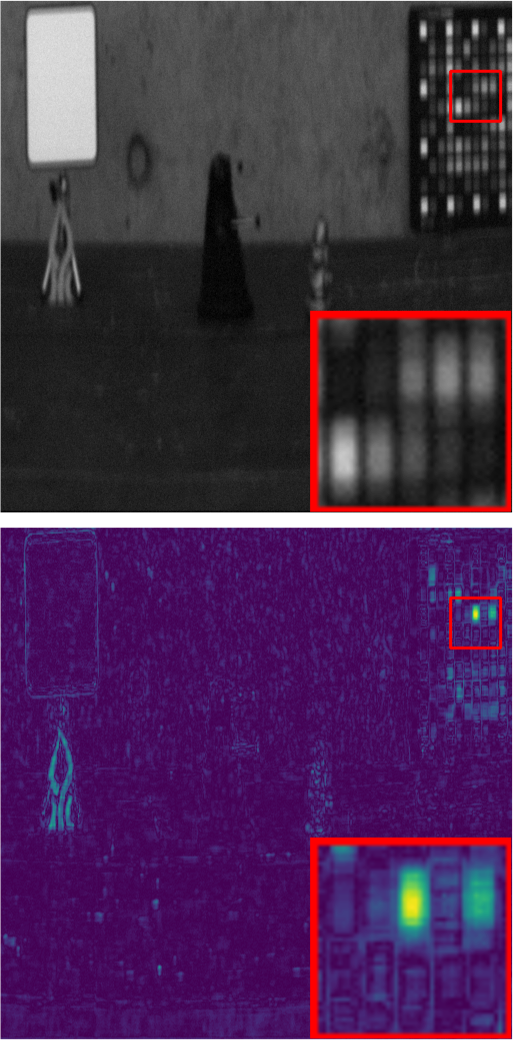}}
         & \multicolumn{1}{m{0.12\linewidth}}{\includegraphics[width=0.85in]{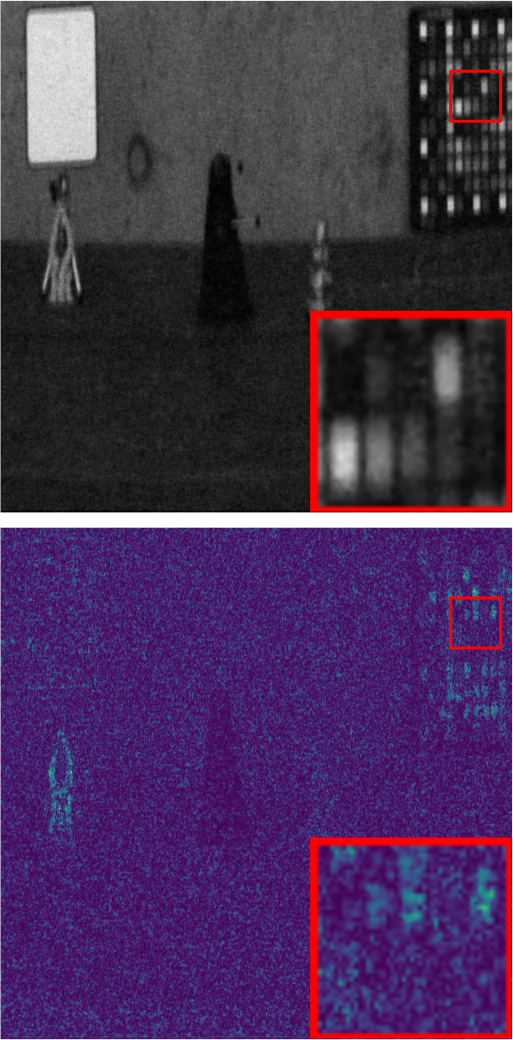}}
         & \multicolumn{1}{m{0.12\linewidth}}{\includegraphics[width=0.85in]{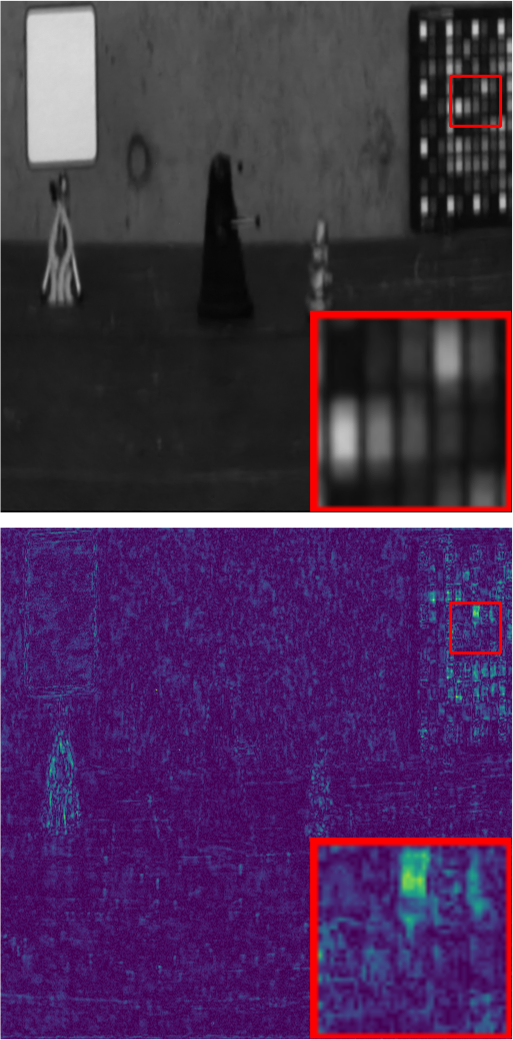}}
         & \multicolumn{1}{m{0.12\linewidth}}{\includegraphics[width=0.85in]{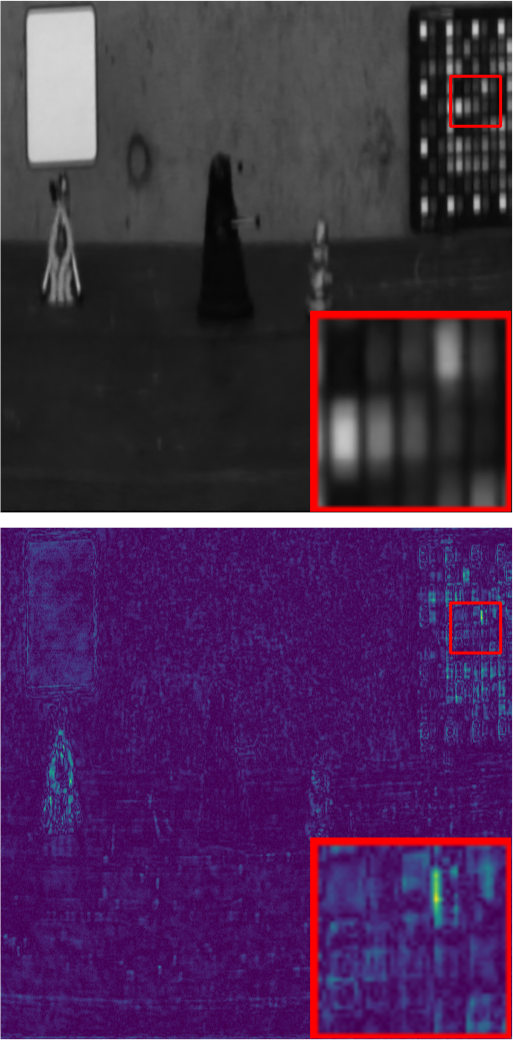}}                                                    \\

        \rotatebox{90}{\quad\quad g+stripe}

         & \multicolumn{1}{m{0.12\linewidth}}{\includegraphics[width=0.85in]{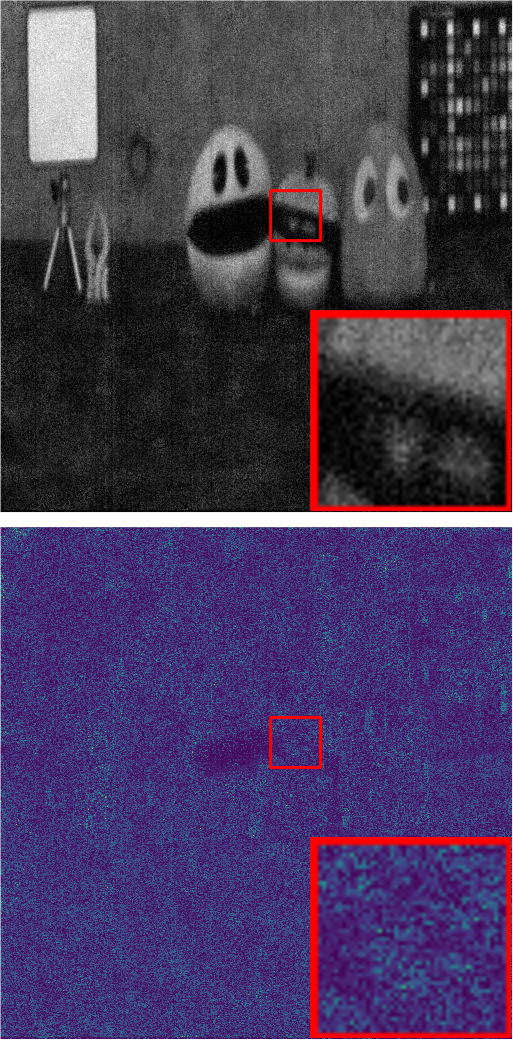}}
         & \multicolumn{1}{m{0.12\linewidth}}{\includegraphics[width=0.85in]{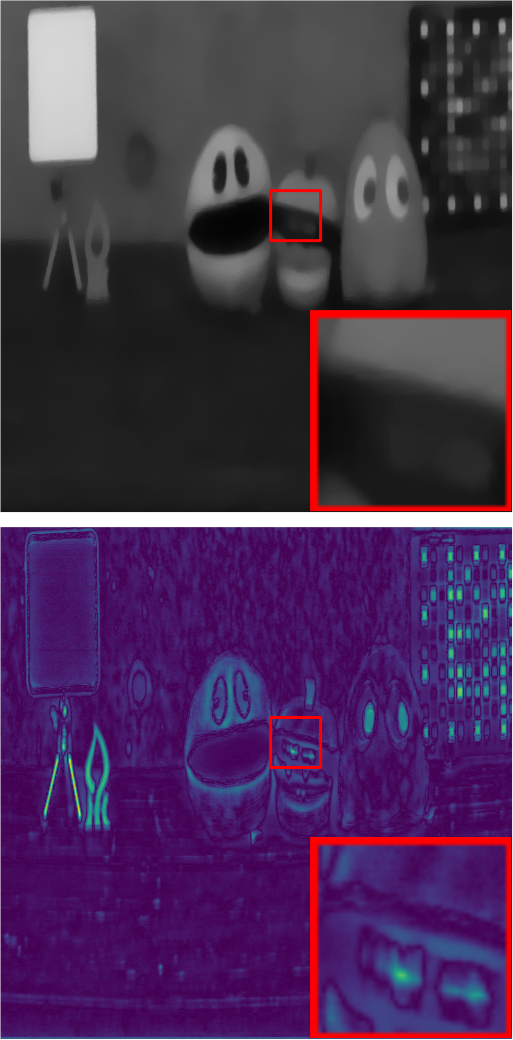}}
         & \multicolumn{1}{m{0.12\linewidth}}{\includegraphics[width=0.85in]{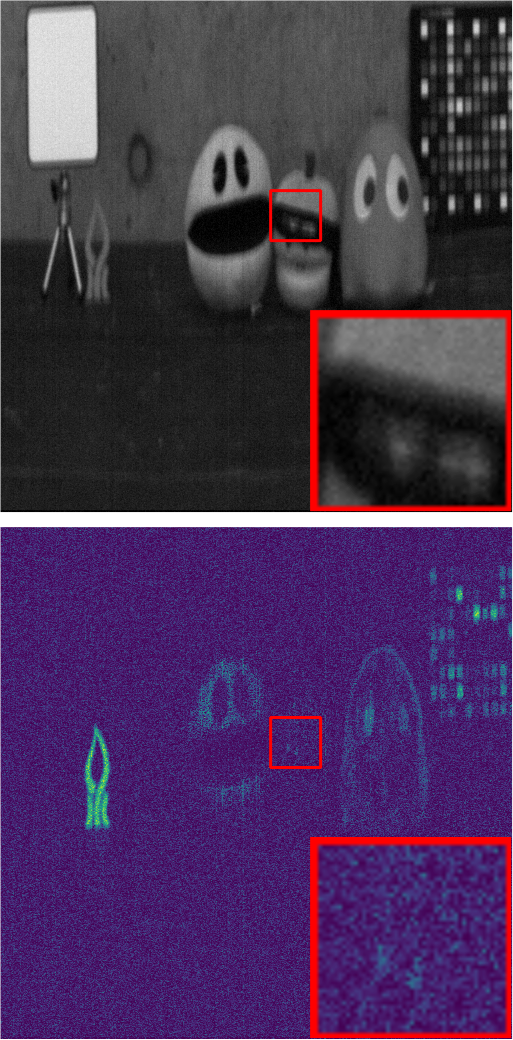}}
         & \multicolumn{1}{m{0.12\linewidth}}{\includegraphics[width=0.85in]{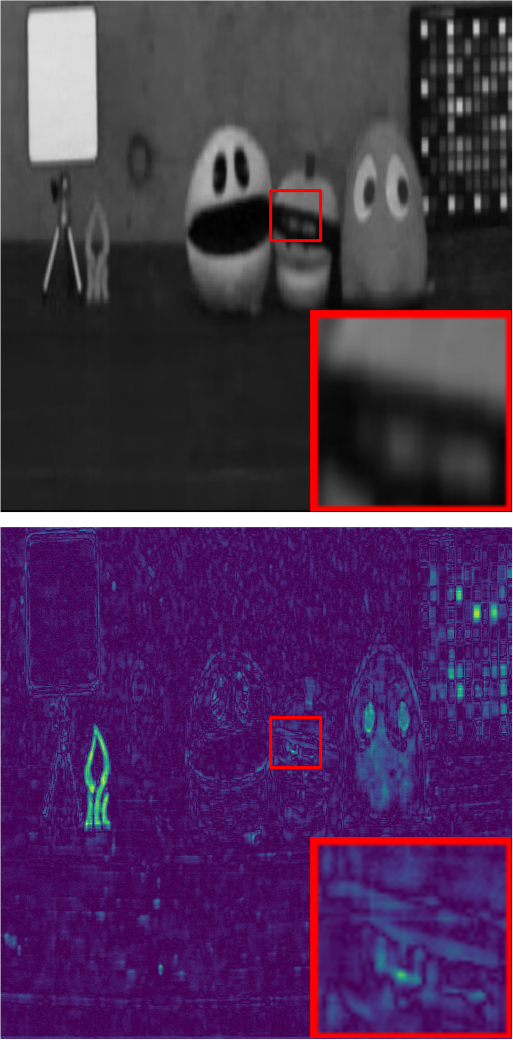}}
         & \multicolumn{1}{m{0.12\linewidth}}{\includegraphics[width=0.85in]{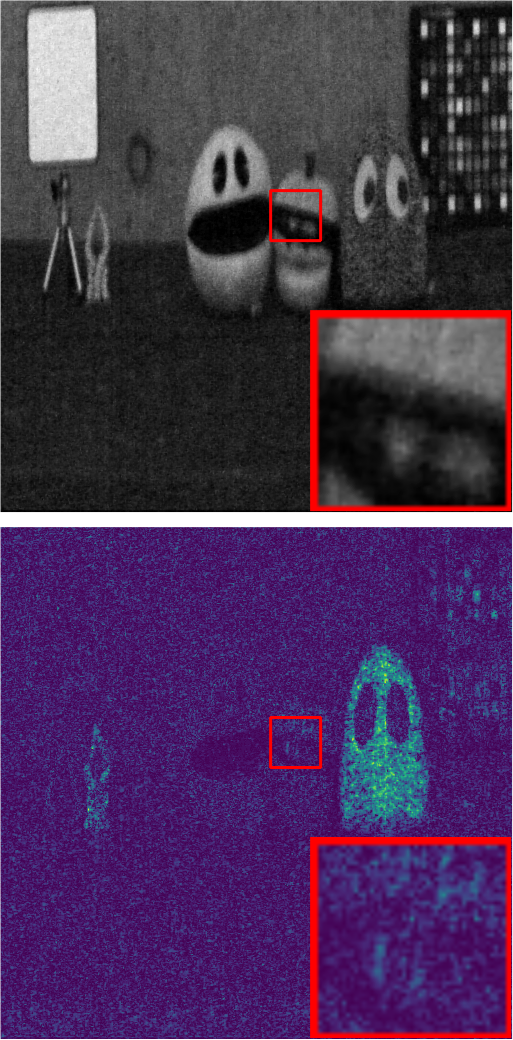}}
         & \multicolumn{1}{m{0.12\linewidth}}{\includegraphics[width=0.85in]{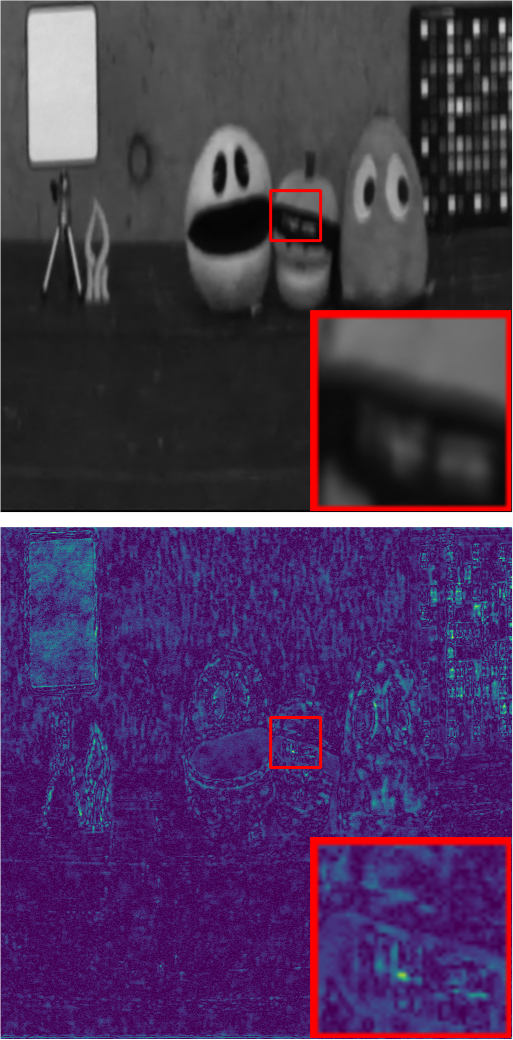}}
         & \multicolumn{1}{m{0.12\linewidth}}{\includegraphics[width=0.85in]{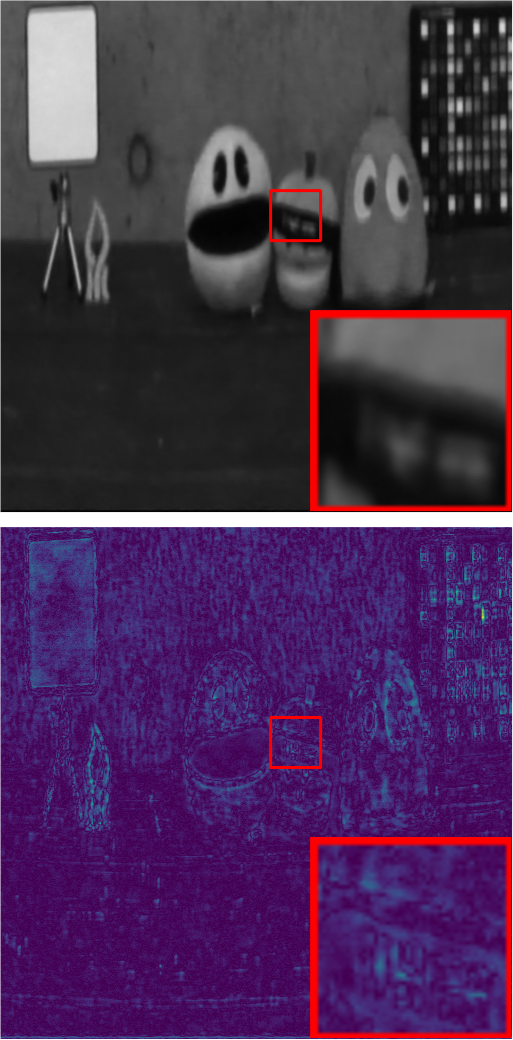}}                                                    \\

        \rotatebox{90}{\quad  g+impulse}

         & \multicolumn{1}{m{0.12\linewidth}}{\includegraphics[width=0.85in]{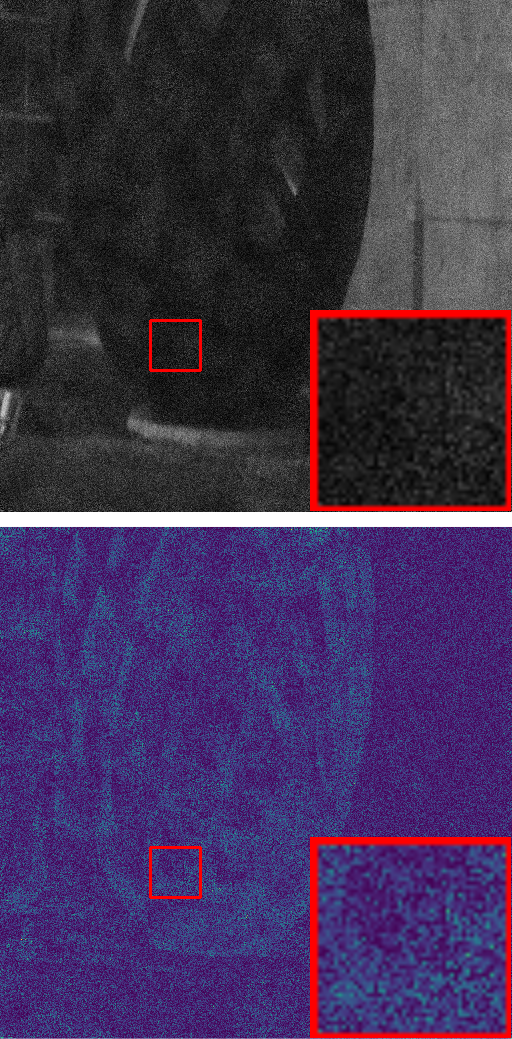}}
         & \multicolumn{1}{m{0.12\linewidth}}{\includegraphics[width=0.85in]{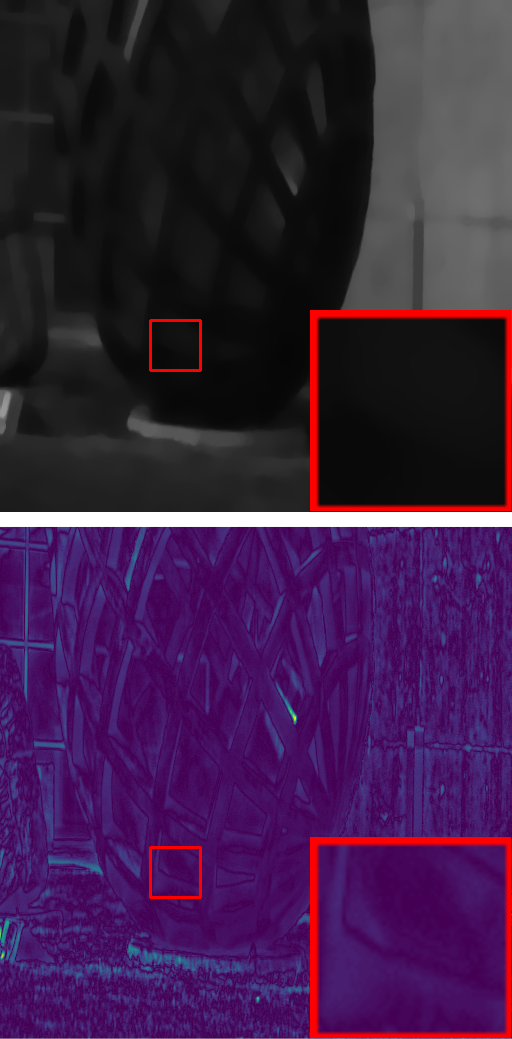}}
         & \multicolumn{1}{m{0.12\linewidth}}{\includegraphics[width=0.85in]{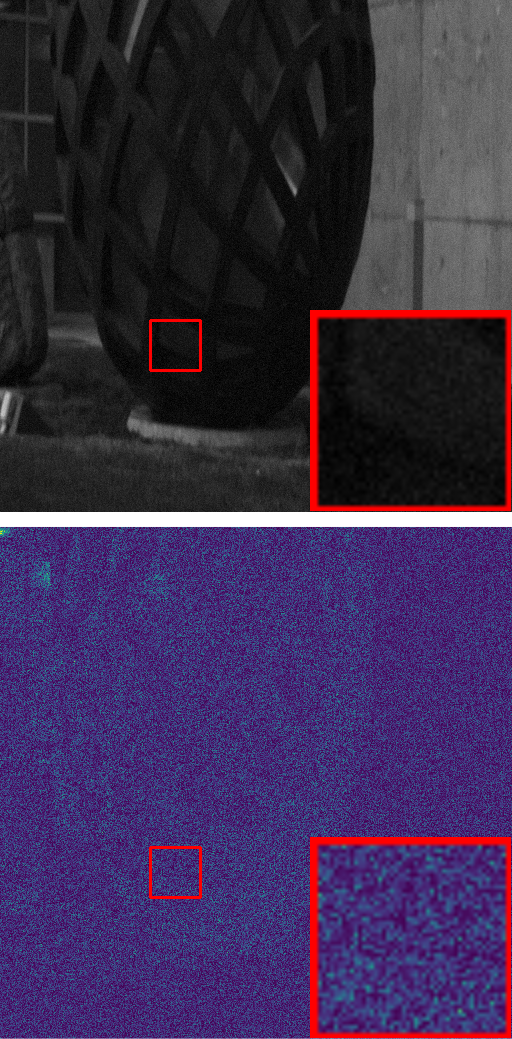}}
         & \multicolumn{1}{m{0.12\linewidth}}{\includegraphics[width=0.85in]{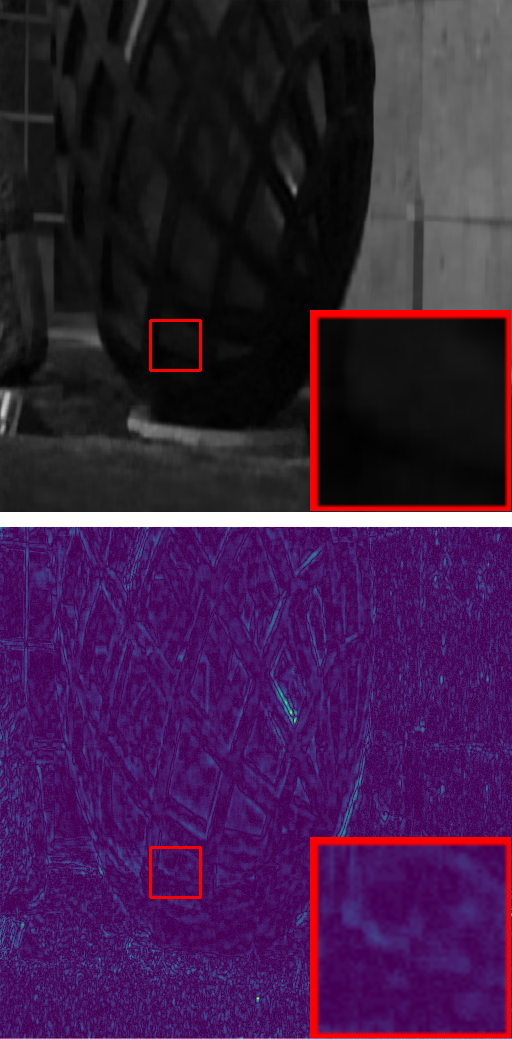}}
         & \multicolumn{1}{m{0.12\linewidth}}{\includegraphics[width=0.85in]{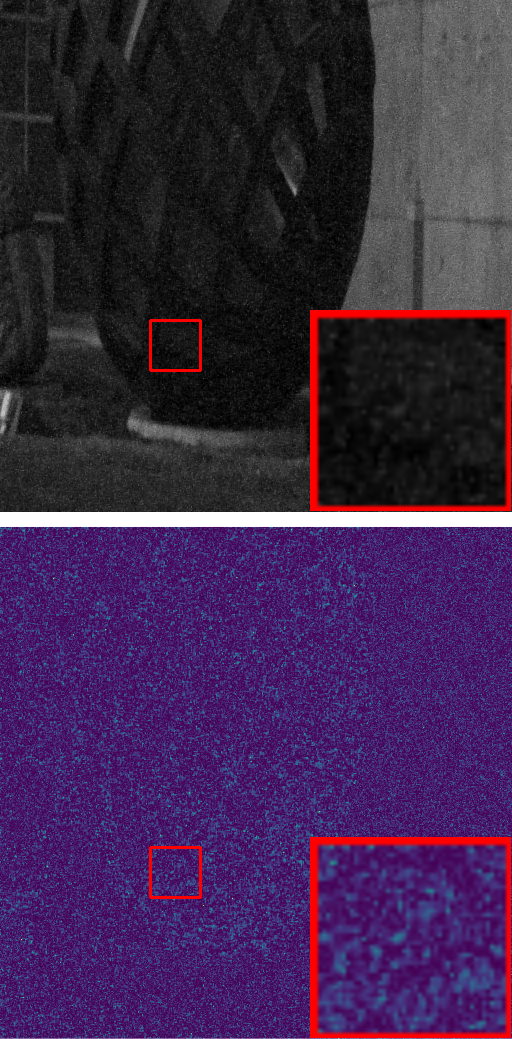}}
         & \multicolumn{1}{m{0.12\linewidth}}{\includegraphics[width=0.85in]{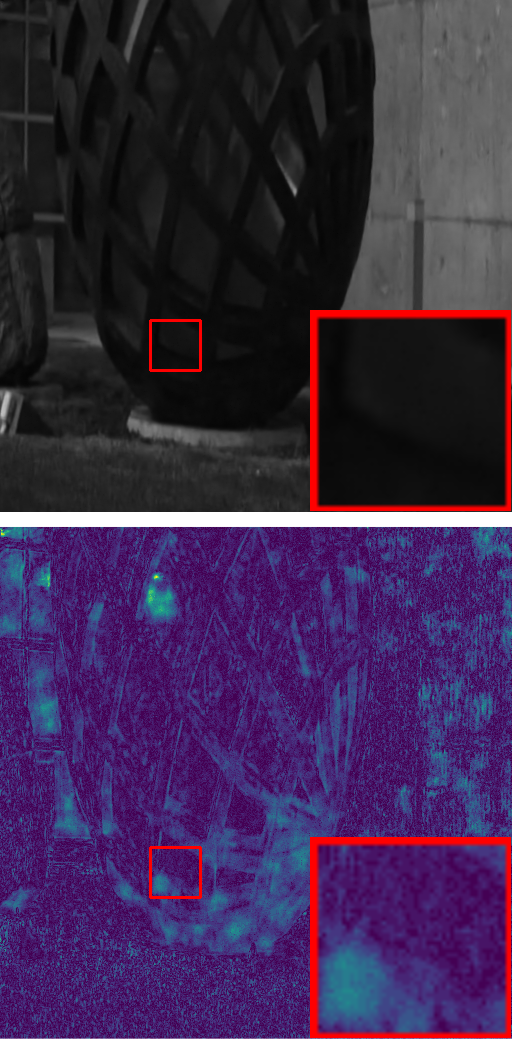}}
         & \multicolumn{1}{m{0.12\linewidth}}{\includegraphics[width=0.85in]{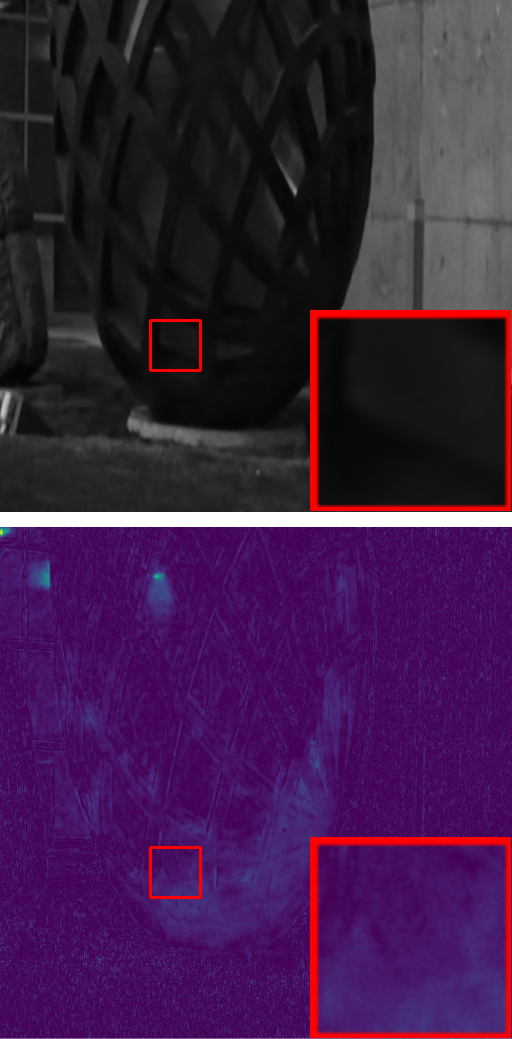}}
    \end{tabular}
    \caption{Simulated Complex noise removal results at the $20^{th}$ band of images on ICVL dataset. The first row shows the denoising results. The second row includes the corresponding error maps.}
    \label{fig:denoise-complex}
\end{figure*}

\subsection{HSI Denoising}\label{sec:denoise}

\subsubsection{Experiments on Gaussian Noise}
As shown in Equation \eqref{eq-admm-v}, a plug-and-play denoiser should be able to tackle Gaussian denoising with a relatively wide range of noise strength. Therefore, we first evaluate the performance of our denoiser on simulated Gaussian noise. Under this setting, additive Gaussian white noise is added to each input HSI with different strengths, including 30, 50, 70, and random strengths ranging from 30 to 70.

We compare our network with five recently developed state-of-the-art HSI denoising algorithms, including three traditional methods, BM4D \cite{maggioni2012nonlocal}, weighted low-Rank tensor recovery (WLRTR) \cite{chang2020weighted} and non-local meets global (NGmeet) \cite{he2019non}, and two deep learning methods, including HSID-CNN \cite{yuan2018hyperspectral} and QRNN3D \cite{wei20203}. For traditional methods, the parameters are manually or automatically set up according to the original papers. For deep learning methods, we fine-tune/retrain their pre-trained models using the same dataset as ours.

The quantitative results under different noise levels for ICVL dataset are summarized in Table \ref{tab:denoise-gaussian}. As we can see, the proposed method outperforms most of the competing methods in terms of PSNR and SSIM. In particular, our method achieves the highest PNSR in the experiment of random noise levels ranging from 30 to 70, which makes it more suitable for plug-and-play denoising. NGmeet is the second-best method and achieves similar performance, but it is far more time-consuming than ours ($\approx$ 100 times). Figure \ref{fig:denoising-gaussian} shows the visual comparison of different methods under noise level $\sigma=50$. As we can see, our method can properly removing the Gaussian noise while preserving the fine-grained details in HSIs. Traditional methods, such as BM4D and WLRTR, introduce evident artifacts in some areas. NGmeet performs better for removing artifacts, but it produces the over-smooth image. The image restored by QRNN3D demonstrates the most similar visual appearance as ours, but it fails to restore a smooth result while facing a large area of white pixels (see the mirror in the upper left corner).

\subsubsection{Experiments on Complex Noise}\label{sec:exp-complex}

To further demonstrate the capability of our denoiser for HSI denoising. We also conduct experiments on simulated complex noise as it is more common in real scenes. Three complex noise cases are tested, including: \emph{non-iid}: each band in the HSI is corrupted by zero-mean Gaussian noise with different intensities ranging from 0 to 70; \emph{g+stripe}: apart from non-iid noise, stripe noise (5\%-15\% percentages of columns) is randomly added to one-third of bands; \emph{g+impulse}: apart from non-iid noise, impulse noise with intensity ranging from 10\% to 70\% is randomly added to one-third of bands.

As the noise level map is not available for complex noise, the proposed network without the noise level map is used for evaluation under these cases. Since traditional methods work best under specific noise assumptions, we compare four traditional baselines different from those used in Gaussian noise. The complete list of competing methods includes low-rank matrix recovery approaches (LRMR \cite{zhang2013hyperspectral}, LRTV \cite{he2015total}, NMoG \cite{chen2017denoising}), low-rank tensor approach (TDTV \cite{wang2017hyperspectral}), and deep-learning methods (HSID-CNN \cite{yuan2018hyperspectral}, QRNN3D \cite{wei20203}).

Table \ref{tab:denoise-commplex} shows the quantitative results under three complex noise cases on ICVL dataset. It can be seen that our method achieves the best performance for all three noise cases. Specifically, our method is significantly better than traditional baselines. QRNN3D achieves the most competitive results as ours, but it is apparently inferior to ours under \emph{g+impulse} noise case. Figure \ref{fig:denoise-complex} provides the visual comparison of our method and competing ones. It can be seen that our method can properly remove the noise while retaining more details than the others. Traditional methods either fail to remove most of the noise (LRMR, NMoG), or produce over-smooth results (LRTV, TTDTV). QRNN3D produces most similar results as ours under \emph{non-iid} and \emph{g+stripe} cases but is worse than ours under \emph{g+impulse} cases, which is consistent with the quantitative results.

\begin{table*}[h]
    \caption{Quantitative super-resolution results of different methods under different scale factors on ICVL dataset. Our method is training-free. \emph{Clean} means the competing methods are trained and tested on clean data. \emph{Noisy} means the competing methods are trained and tested on noisy data. \emph{Clean2Noisy} means the competing methods are trained on clean data but tested on noisy one. }
    \begin{center}
        \begin{tabular}{|c|c|c|c|c|c|c|c|c|c|}
            \hline
            \multirow{2}{*}{Method}        & \multicolumn{3}{c|}{Clean} & \multicolumn{3}{c|}{Noisy} & \multicolumn{3}{c|}{Clean2Noisy}                                                                                                               \\ \cline{2-10}
                                           & PSNR                       & SSIM                       & SAM                              & PSNR             & SSIM            & SAM             & PSNR             & SSIM            & SAM             \\ \hline
            \multicolumn{10}{|c|}{Gaussian ($8\times8$, $\sigma=3$); Scale Factor = 2}                                                                                                                                                                \\ \hline
            Bicubic                        & 35.1375                    & 0.9575                     & 0.0301                           & 29.2433          & 0.8139          & 0.3050          & 29.2433          & 0.8139          & 0.3050          \\ \hline
            3D-FCNN\cite{mei20173dfcnn}    & 39.8415                    & 0.9790                     & 0.0295                           & 36.7468          & 0.9626          & 0.0849          & 26.3227          & 0.6667          & 0.4084          \\ \hline
            SSPSR\cite{jiang2020learning}  & 47.5592                    & 0.9955                     & 0.0230                           & 37.9793          & \textbf{0.9752} & 0.0758          & 11.9596          & 0.0706          & 1.0787          \\ \hline
            IFN\cite{hu2020hyperspectral}  & 41.0717                    & 0.9844                     & 0.0279                           & 35.7368          & 0.9531          & 0.1023          & 20.3987          & 0.3404          & 0.7170          \\ \hline
            Bi-3DQRNN\cite{fu2021biqrnn3d} & 42.5320                    & 0.9891                     & 0.0288                           & 38.1155          & 0.9726          & 0.0660          & 22.7233          & 0.5232          & 0.5604          \\ \hline
            Ours                           & \textbf{48.7561}           & \textbf{0.9956}            & \textbf{0.0200}                  & \textbf{38.1213} & 0.9728          & \textbf{0.059}  & \textbf{38.1213} & \textbf{0.9728} & \textbf{0.059}  \\ \hline
            \multicolumn{10}{|c|}{Gaussian ($8\times8$, $\sigma=3$); Scale Factor = 4}                                                                                                                                                                \\ \hline
            Bicubic                        & 35.1230                    & 0.9543                     & 0.0309                           & 28.5231          & 0.8006          & 0.3303          & 28.5231          & 0.8006          & 0.3303          \\ \hline
            3D-FCNN\cite{mei20173dfcnn}    & 38.4379                    & 0.9729                     & 0.0309                           & 35.1536          & 0.9506          & 0.1052          & 26.7587          & 0.7140          & 0.4003          \\ \hline
            SSPSR\cite{jiang2020learning}  & 39.1961                    & 0.9798                     & 0.0355                           & \textbf{36.6960} & \textbf{0.9670} & \textbf{0.0658} & 22.1606          & 0.4445          & 0.5990          \\ \hline
            IFN\cite{hu2020hyperspectral}  & 38.9177                    & 0.9749                     & 0.0297                           & 33.7173          & 0.9341          & 0.1338          & 21.0832          & 0.3852          & 0.6851          \\ \hline
            Bi-3DQRNN\cite{fu2021biqrnn3d} & 39.5634                    & 0.9794                     & 0.0303                           & 36.5558          & 0.9634          & 0.0716          & 24.9716          & 0.5828          & 0.5359          \\ \hline
            Ours                           & \textbf{40.9562}           & \textbf{0.9808}            & \textbf{0.0253}                  & 36.5483          & 0.9601          & 0.0794          & \textbf{36.5483} & \textbf{0.9601} & \textbf{0.0794} \\ \hline
            \multicolumn{10}{|c|}{Gaussian ($8\times8$, $\sigma=3$); Scale Factor = 8}                                                                                                                                                                \\ \hline
            Bicubic                        & 32.3452                    & 0.9262                     & 0.0387                           & 27.3477          & 0.8018          & 0.3382          & 27.3477          & 0.8018          & 0.3382          \\ \hline
            3D-FCNN\cite{mei20173dfcnn}    & 32.9615                    & 0.9324                     & 0.0391                           & 31.4878          & 0.9174          & 0.1205          & 27.3707          & 0.8040          & 0.3451          \\ \hline
            SSPSR\cite{jiang2020learning}  & 32.4483                    & 0.9336                     & 0.0499                           & 31.8391          & \textbf{0.9268} & 0.0867          & 27.1837          & 0.7553          & 0.3669          \\ \hline
            IFN\cite{hu2020hyperspectral}  & 33.0492                    & 0.9326                     & 0.0380                           & 30.4145          & 0.8995          & 0.1592          & 26.0573          & 0.7121          & 0.4347          \\ \hline
            Bi-3DQRNN\cite{fu2021biqrnn3d} & 33.0909                    & 0.9365                     & 0.0388                           & 32.0505          & 0.9251          & 0.0887          & 24.6905          & 0.7376          & 0.4121          \\ \hline
            Ours                           & \textbf{33.5676}           & \textbf{0.9367}            & \textbf{0.0375}                  & \textbf{32.5095} & 0.9251          & \textbf{0.0774} & \textbf{32.5095} & \textbf{0.9272} & \textbf{0.0774} \\ \hline
        \end{tabular}
    \end{center}

    \label{tab:super-resolution}
\end{table*}

\begin{figure*}[h!]

    \centering
    \subfigure[Bicubic]{
        \includegraphics[width=1in]{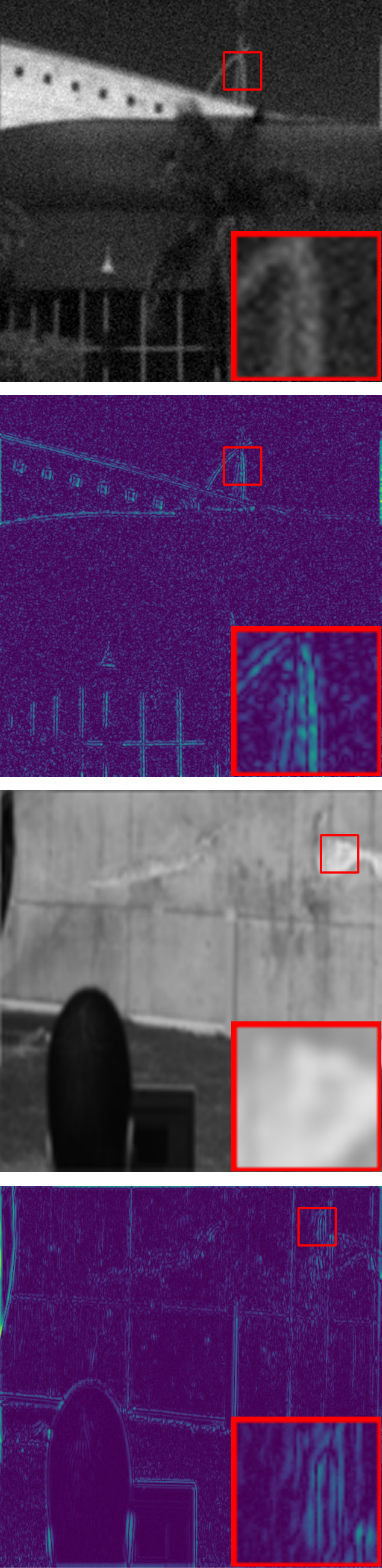}
    }
    \subfigure[IFN]{
        \includegraphics[width=1in]{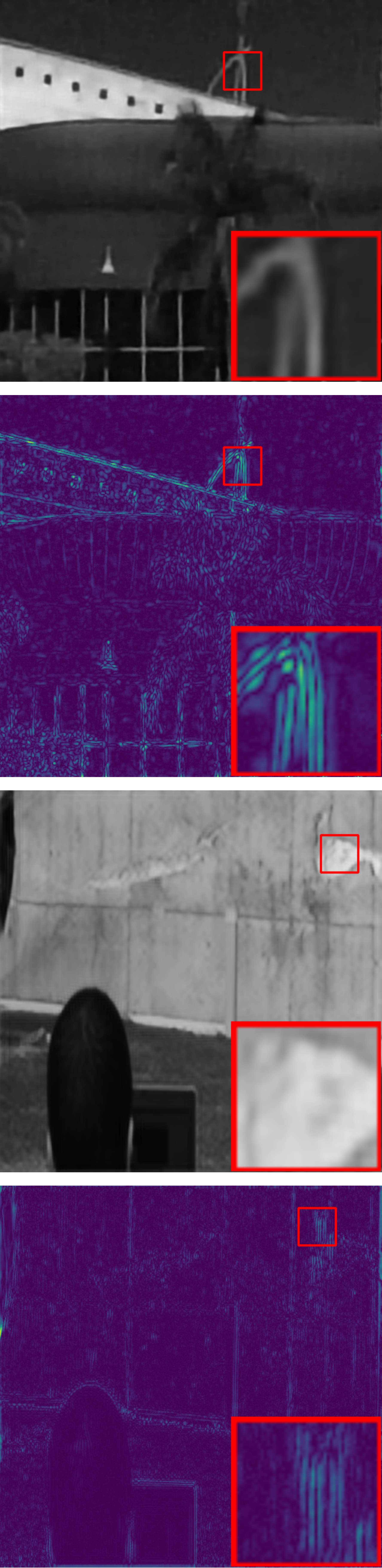}
    }
    \subfigure[3D-FCNN]{
        \includegraphics[width=1in]{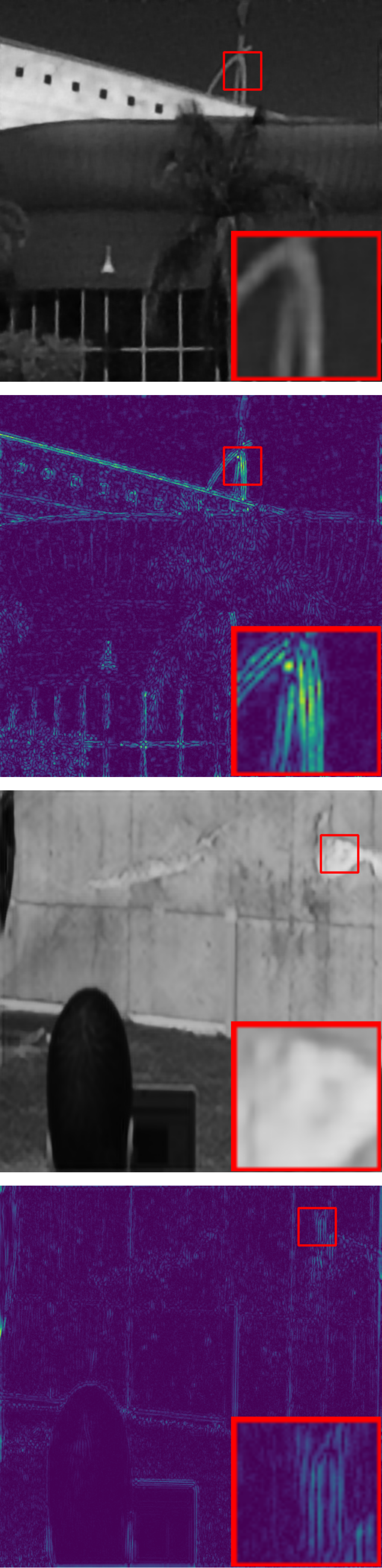}
    }
    \subfigure[SSPSR]{
        \includegraphics[width=1in]{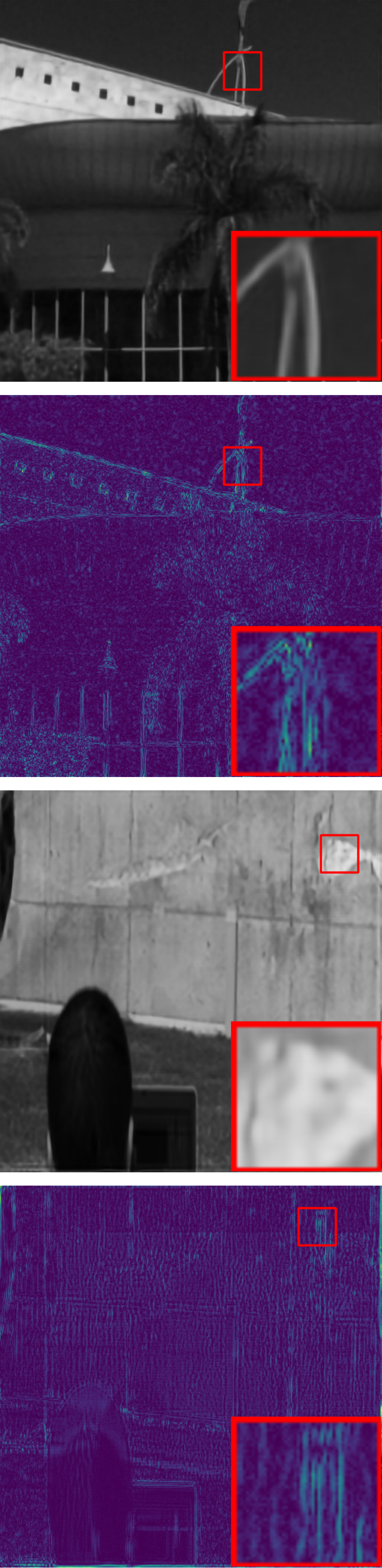}
    }
    \subfigure[Bi-QRNN3D]{
        \includegraphics[width=1in]{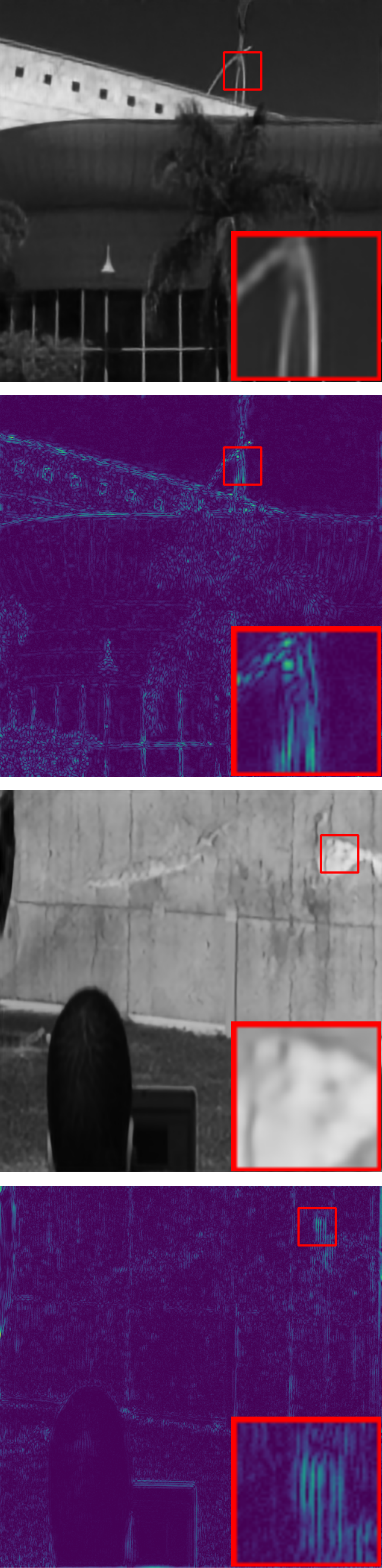}
    }
    \subfigure[Ours]{
        \includegraphics[width=1in]{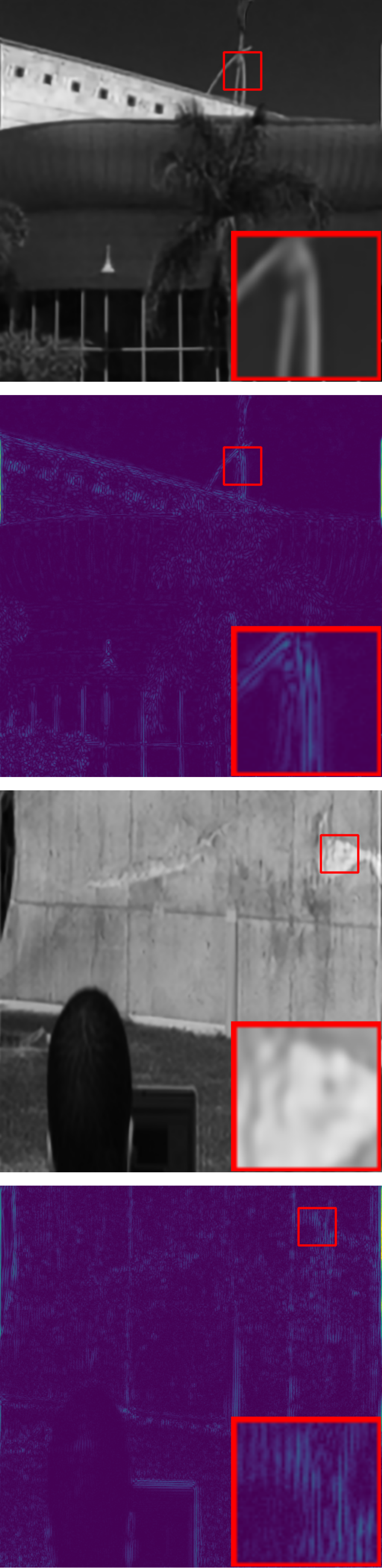}
    }
    \caption{The super-resolution results at the $20^{th}$ band of images on the ICVL dataset for \emph{Noisy} (row 1-2, scale factor=2) and \emph{Clean} (row 2-3, scale factor=4) cases. The first row shows the super-resolution results. The second row includes the corresponding error maps. }
    \label{fig:super-resolution-result}
\end{figure*}

\subsection{HSI Super-Resolution}\label{sec:sr}
Without any extra training, our denoiser can be plugged into the PnP-ADMM algorithm to solve the HSI super-resolution problem. To demonstrate the flexibility and powerfulness of our method, we test our method under \emph{Clean}, \emph{Noisy}, and \emph{Clean2Noisy} settings for three scale factors, \ie, 2, 4, 8. Specifically, the low-resolution HSIs are generated by first blurring the original HSIs via a $8\times8$ Gaussian kernel with $\sigma=3$, and then downsampled by a scaling factor of 2, 4, or 8. Under \emph{Noisy} and \emph{Clean2Noisy} settings, additional Gaussian noise with $\sigma_n=10$ is added.

For comparison, we adopt the bicubic interpolation as the baseline and compare our method against four deep-learning-based methods, \ie, IFN \cite{hu2020hyperspectral}, 3D-FCNN \cite{mei20173dfcnn}, SSPSR \cite{jiang2020learning}, and Bi-3DQRNN \cite{fu2021biqrnn3d}. Traditional HSI super-resolution methods \cite{dian2020regularizing, dian2019learning, fu2019hyperspectral} are not considered as most of them focus on super-resolution with additional inputs, such as multiple frames and RGB images. We directly apply our PnP framework to the aforementioned settings without any extra training, while all the competing methods are trained with the corresponding datasets for \emph{Clean} and \emph{Noisy} settings. In the setting of \emph{Clean2Noisy}, the models from \emph{Clean} settings are directly applied to evaluate the generalizability of each method.

The quantitative results of comparison on ICVL dataset are provided in Table \ref{tab:super-resolution}. As we can see, our method obtains the best results under \emph{Clean} and \emph{Clean2Noisy} settings. Under \emph{Noisy} setting, our method achieves better results against IFN and 3D-FCNN and achieves the best PSNR at the scale factors of $\times 2$ and $\times 8$. SSPSR is the second-best method and achieves better results than ours when scale factor = 2 under \emph{Noisy} setting. It can be partially explained that SSPSR is specifically trained on the noisy dataset, while ours is training-free. Nevertheless, when it comes to \emph{Clean2Noisy} setting, SSPSR trained on the clean dataset completely fails to generalize to the noisy testset, as all the other competing methods do. As for visual comparisons shown in Figure \ref{fig:super-resolution-result}, we can observe that the images produced by our method retain most details for \emph{Clean} case. As for \emph{Noisy} case, ours is prominently better than IFN and 3DFCNN but a little more blurred than SSPSR. Overall, our method achieves the competitive or even better performance against the deep-learning-based methods \textit{without any task-specific training}.

\subsection{HSI Compressed Sensing}\label{sec:cs}

Compressive HSI imaging is a common technique for improving the spatial and temporal resolution of HSI. Such a system captures a single snapshot measurement frame that encodes the information that can be used for recovering the original high-dimensional data with specific algorithms.  In this experiment, we extend our method for coded aperture snapshot spectral imagers (CASSI \cite{wagadarikar2009video}). Three state-of-the-art compressed HSI reconstruction methods for CASSI are chosen for comparison, including GAP-TV \cite{yuan2016generalized}, DeSCI \cite{liu2018rank} and NGmeet \cite{he2020nonlocal}. Following the setting in \cite{liu2018rank}, the CAVE Toy image is used for the experiment and the simulated data is generated by the same compressed operator.

\begin{table}[h!]
    \caption{Quantitative compressed HSI reconstruction results of different methods on the CAVE Toy image.}
    \label{tab:cs}
    \begin{center}
        \small
        \begin{tabular}{@{}lllll@{}}
            \toprule
            Method & GAP-TV\cite{yuan2016generalized} & DeSCI\cite{liu2018rank} & NGmeet\cite{he2020nonlocal} & Ours           \\ \midrule
            PSNR   & 24.66                            & 25.91                   & 27.13                       & \textbf{30.56} \\
            SSIM   & 0.861                            & 0.909                   & 0.913                       & \textbf{0.945} \\ \bottomrule
        \end{tabular}
    \end{center}
\end{table}

The reconstruction results of GAP-TV \cite{yuan2016generalized} and DeSCI \cite{liu2018rank} were provided in \cite{liu2018rank}, and the results of NGmeet were provided in \cite{he2020nonlocal}. Table \ref{tab:cs} shows the quantitative results of our method and the competing ones. It can be seen that our method is significantly better in the metrics of both PSNR and SSIM.

\subsection{HSI Inpainting}\label{sec:inpainting}

\begin{figure*}[h]
    \centering
    \subfigure[Degraded image ]{
        \includegraphics[width=0.18\linewidth]{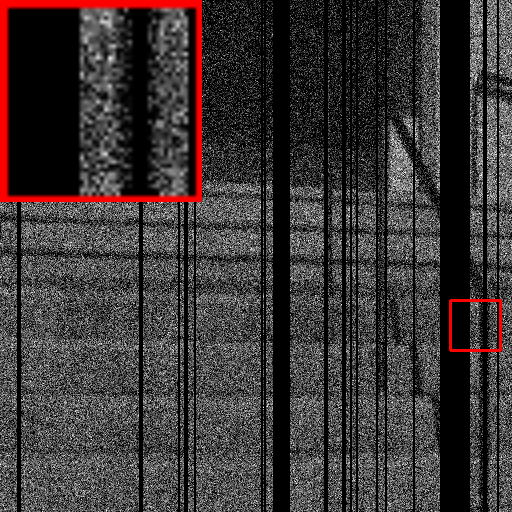}
    }
    \subfigure[Ground truth]{
        \includegraphics[width=0.18\linewidth]{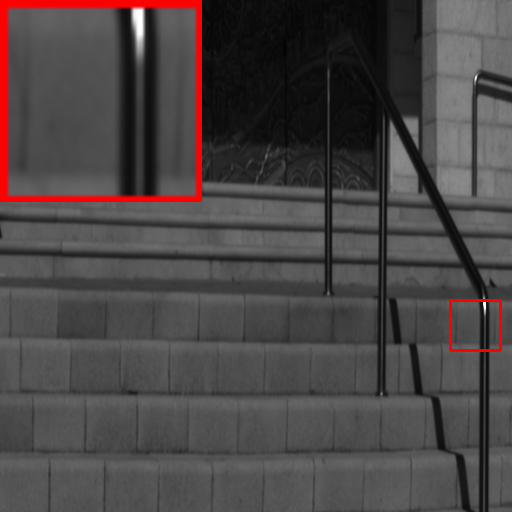}
    }
    \subfigure[WLRTR]{
        \includegraphics[width=0.18\linewidth]{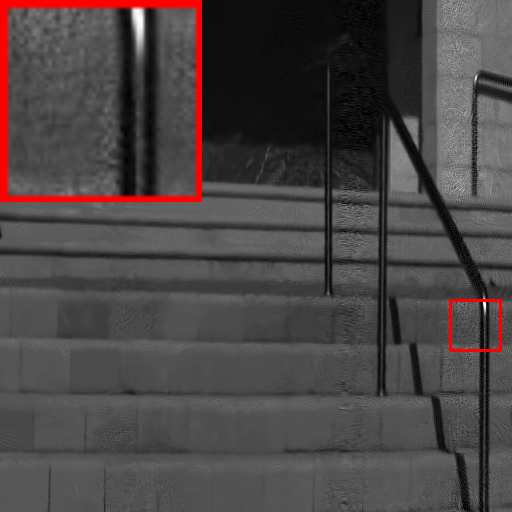}
    }
    \subfigure[FastHyIn]{
        \includegraphics[width=0.18\linewidth]{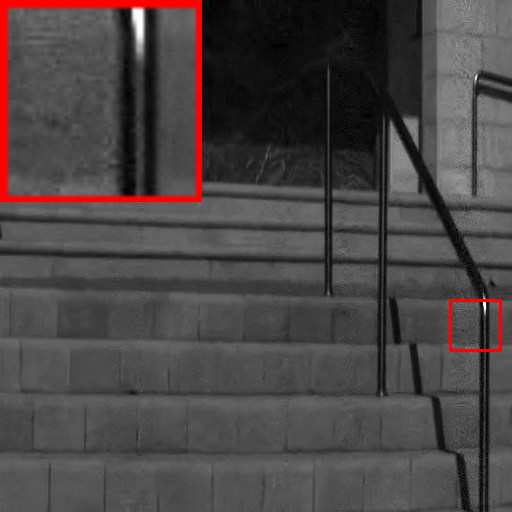}
    }
    \subfigure[Ours]{
        \includegraphics[width=0.18\linewidth]{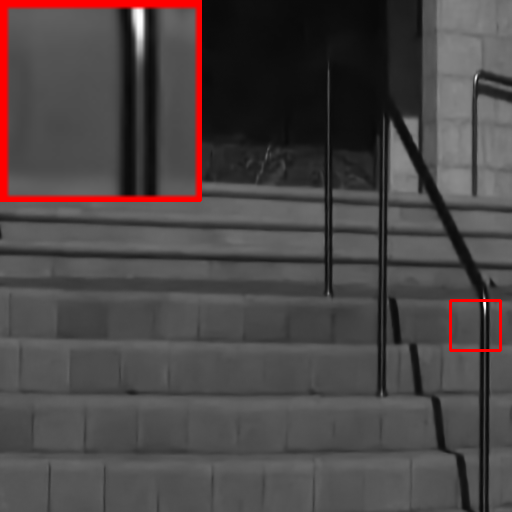}
    }
    \caption{Simulated inpainting results under stripe mask and Gaussian noise ($\sigma=50$) on ICVL dataset.}
    \label{fig:inpainting-result}
\end{figure*}

We also extend our method for HSI inpainting and compare the performance against two recent proposed HSI inpainting methods, \ie, weighted low-rank tensor recovery (WLRTR) \cite{chang2020weighted}, and FastHyIn \cite{zhuang2018fast}. We generate the simulated data with stripe masks where stripes are randomly distributed in all bands. To simulate the real degraded HSIs, the Gaussian noise of two noise levels 30 and 50 are added.

As shown in Figure  \ref{fig:inpainting-result}, although WLRTR and FastHyIn can remove the extra Gaussian noise properly, both of them fail to remove the stripes completely. In contrast, our method produces better results with diminished stripes and noise. The qualitative assessment results are listed in Table \ref{tab:destripe}. Compared with all competing methods, our plug-and-play method achieves the best performance in all qualitative/quantitative assessments, further confirming the high fidelity of our method.

\begin{table}[h]
    \caption{Quantitative inpainting results of different methods under stripe mask and different noise levels on ICVL dataset.}

    \begin{center}
        \small
        \begin{tabular}{@{}ccccc@{}}
            \toprule
            Method & Noisy & WLRTR\cite{chang2020weighted} & FastHyIn\cite{zhuang2018fast} & Ours           \\ \midrule
            \multicolumn{5}{c}{Noise level = 30}                                                            \\ \midrule
            PSNR   & 17.21 & 41.90                         & 39.10                         & \textbf{42.43} \\
            SSIM   & 0.234 & 0.988                         & 0.976                         & \textbf{0.987} \\
            SAM    & 0.879 & 0.076                         & 0.130                         & \textbf{0.055} \\ \midrule
            \multicolumn{5}{c}{Noise level = 50}                                                            \\ \midrule
            PSNR   & 13.97 & 38.11                         & 36.25                         & \textbf{40.26} \\
            SSIM   & 0.106 & 0.969                         & 0.958                         & \textbf{0.982} \\
            SAM    & 1.000 & 0.101                         & 0.173                         & \textbf{0.063} \\ \bottomrule
        \end{tabular}
    \end{center}

    \label{tab:destripe}
\end{table}

\subsection{Discussion and Analysis} \label{sec:abl}

In this section, we further discuss and analyze the proposed method. We first demonstrate the functionality of each network component in our deep denoiser, whose effectiveness is further verified by comparing it with two traditional Gray/RGB PnP denoisers. Then, we present the inpainting results of our method on the other three unseen HSI datasets, \ie, CAVE \cite{CAVE_0293}, Harvard \cite{harvard}, and Pavia University \cite{paviaU}, to evaluate the generalizability of our method. At the end of this section, we analyze the computational complexity of the proposed method and provide comparisons against other methods.

\subsubsection{Ablation Study on the Proposed Denoiser}

To verify the effectiveness of the proposed components in our deep denoiser, \ie residual block, gated recurrent convolution, and additional noise level map input, we compare the performance of the complete network and the variants that are generated by gradually removing the aforementioned components one by one. All networks are trained with the same dataset and training strategy. Specifically, we denote the complete network as \emph{Full}, the network without noise level map as \emph{-Map}, the network without both noise level map and residual connection as \emph{-Res}, and the network without all the three components as \emph{-Gate}. The quantitative results of these variants on ICVL dataset are listed in Table \ref{tab:denoise-aba}. Take noise level 30 as an example, it can be easily observed that there is a significant improvement (1.84dB) after adding the gated recurrent convolution, which demonstrates the importance of exploiting the spectral correlation in HSIs. Besides, there are also prominent improvements after adding residual block and noise level map, \ie 0.83dB and 0.52dB, which verifies the corresponding effectiveness.

\begin{table}[h!]
    \caption{Ablation study of the proposed denoising network on ICVL dataset.}
    \small
    \begin{center}
        \begin{tabular}{@{}cccccc@{}}
            \toprule
            \multirow{2}{*}{Sigma}    & \multicolumn{1}{c}{\multirow{2}{*}{Metric}} & \multicolumn{4}{c}{Method}                                    \\ \cmidrule(l){3-6}
                                      & \multicolumn{1}{c}{}                        & $-$Gate                    & $-$Res & $-$Map & Full           \\ \midrule
            \multirow{3}{*}{30}       & PSNR                                        & 39.87                      & 41.71  & 42.54  & \textbf{43.06} \\
                                      & SSIM                                        & 0.979                      & 0.987  & 0.989  & \textbf{0.990} \\
                                      & SAM                                         & 0.083                      & 0.059  & 0.054  & \textbf{0.051} \\ \bottomrule
            \multirow{3}{*}{50}       & PSNR                                        & 38.65                      & 39.83  & 40.52  & \textbf{40.92} \\
                                      & SSIM                                        & 0.974                      & 0.980  & 0.983  & \textbf{0.984} \\
                                      & SAM                                         & 0.092                      & 0.073  & 0.067  & \textbf{0.060} \\ \bottomrule
            \multirow{3}{*}{[30, 70]} & PSNR                                        & 39.51                      & 40.98  & 41.81  & \textbf{42.24} \\
                                      & SSIM                                        & 0.977                      & 0.983  & 0.986  & \textbf{0.987} \\
                                      & SAM                                         & 0.087                      & 0.065  & 0.060  & \textbf{0.057} \\ \bottomrule
        \end{tabular}
    \end{center}
    \label{tab:denoise-aba}
\end{table}

\begin{figure*}[h]
    \centering
    \setlength{\tabcolsep}{0.09cm}
    \begin{tabular}{cccccc}
         & GT                                                                                                                    & Bicubic & IRCNN & DRUnet & Ours \\
        \rotatebox{90}{$\times 2$}
         & \multicolumn{1}{m{0.18\linewidth}}{\includegraphics[width=1.2in]{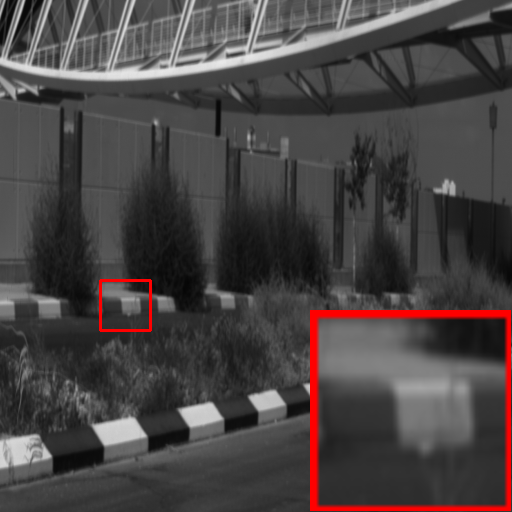}}
         & \multicolumn{1}{m{0.18\linewidth}}{\includegraphics[width=1.2in]{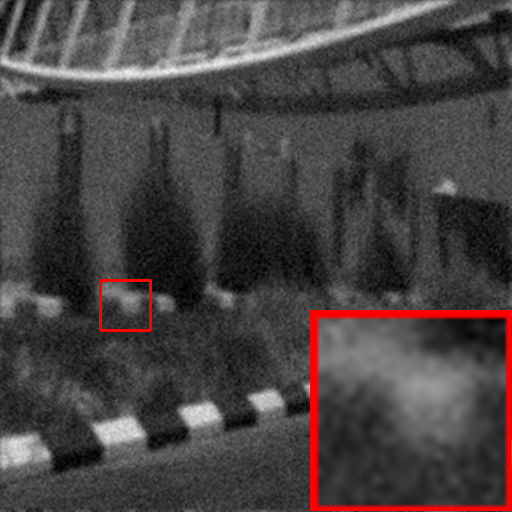}}
         & \multicolumn{1}{m{0.18\linewidth}}{\includegraphics[width=1.2in]{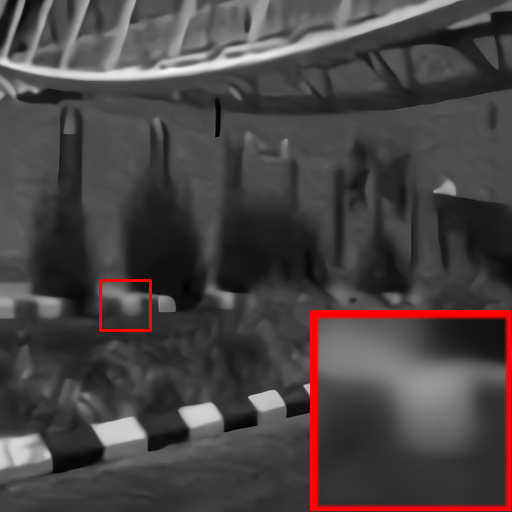}}
         & \multicolumn{1}{m{0.18\linewidth}}{\includegraphics[width=1.2in]{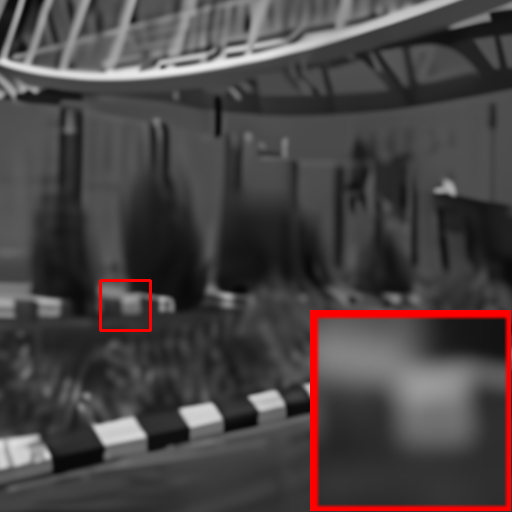}}
         & \multicolumn{1}{m{0.18\linewidth}}{\includegraphics[width=1.2in]{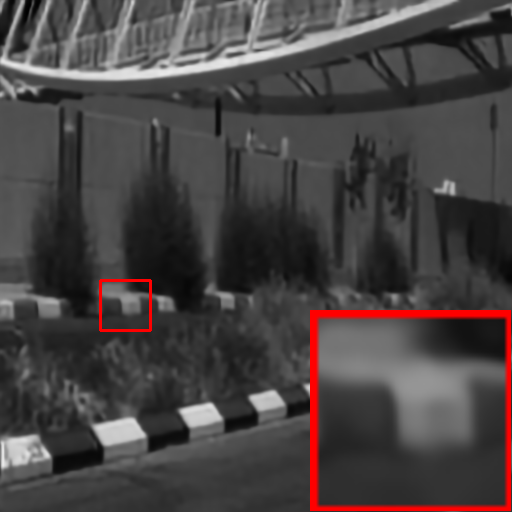}}                                         \\

        \rotatebox{90}{$\times 8$}
         & \multicolumn{1}{m{0.18\linewidth}}{\includegraphics[width=1.2in]{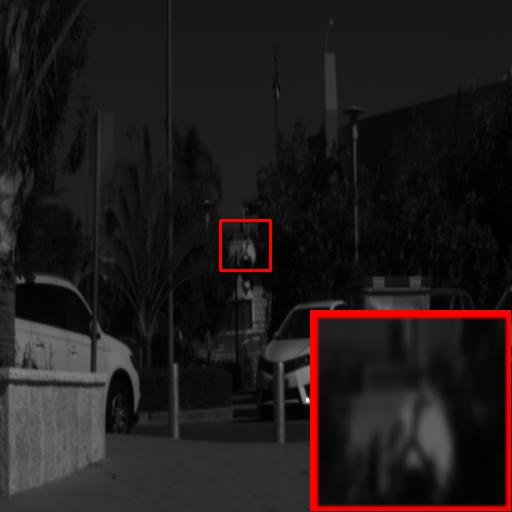}}
         & \multicolumn{1}{m{0.18\linewidth}}{\includegraphics[width=1.2in]{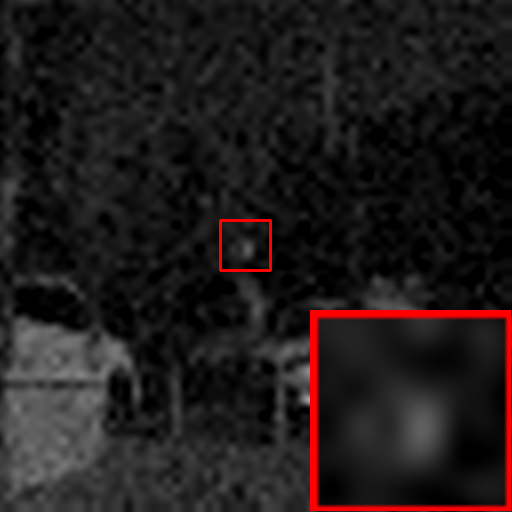}}
         & \multicolumn{1}{m{0.18\linewidth}}{\includegraphics[width=1.2in]{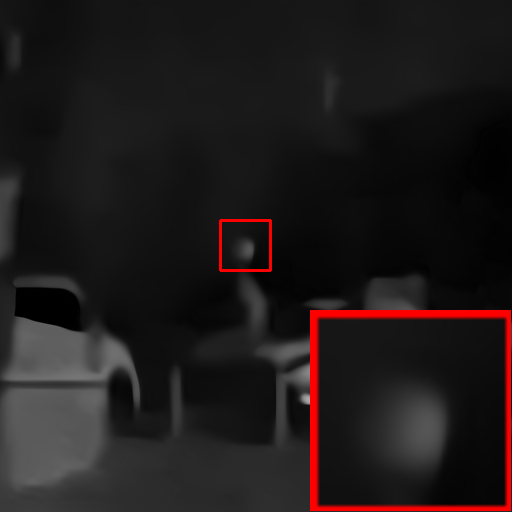}}
         & \multicolumn{1}{m{0.18\linewidth}}{\includegraphics[width=1.2in]{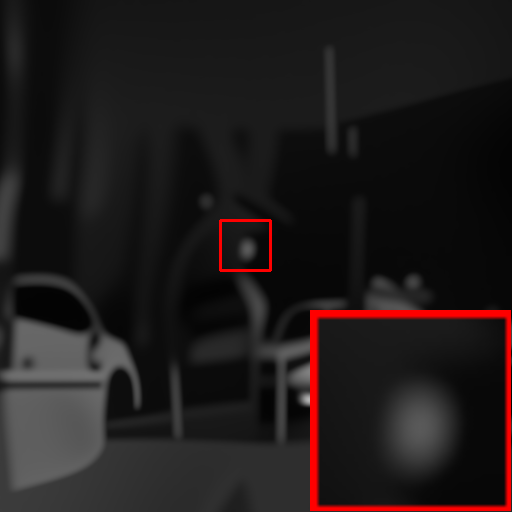}}
         & \multicolumn{1}{m{0.18\linewidth}}{\includegraphics[width=1.2in]{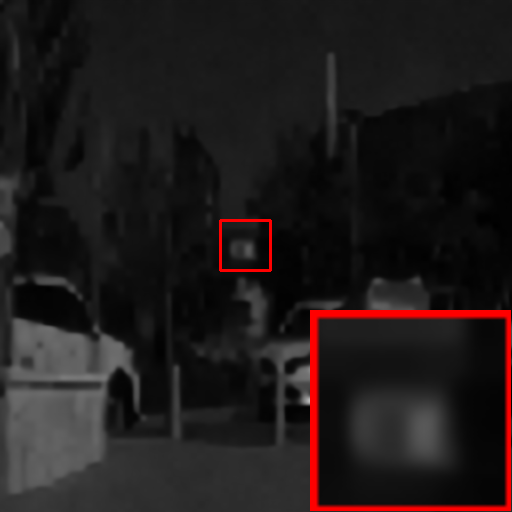}}
    \end{tabular}
    \caption{Visual comparison with traditional PnP denoisers on super-resolution task and ICVL dataset.}
    \label{fig:cmp2d-result}
\end{figure*}

\begin{figure*}[ht!]
    \centering
    \subfigure{
        \includegraphics[width=0.48\linewidth]{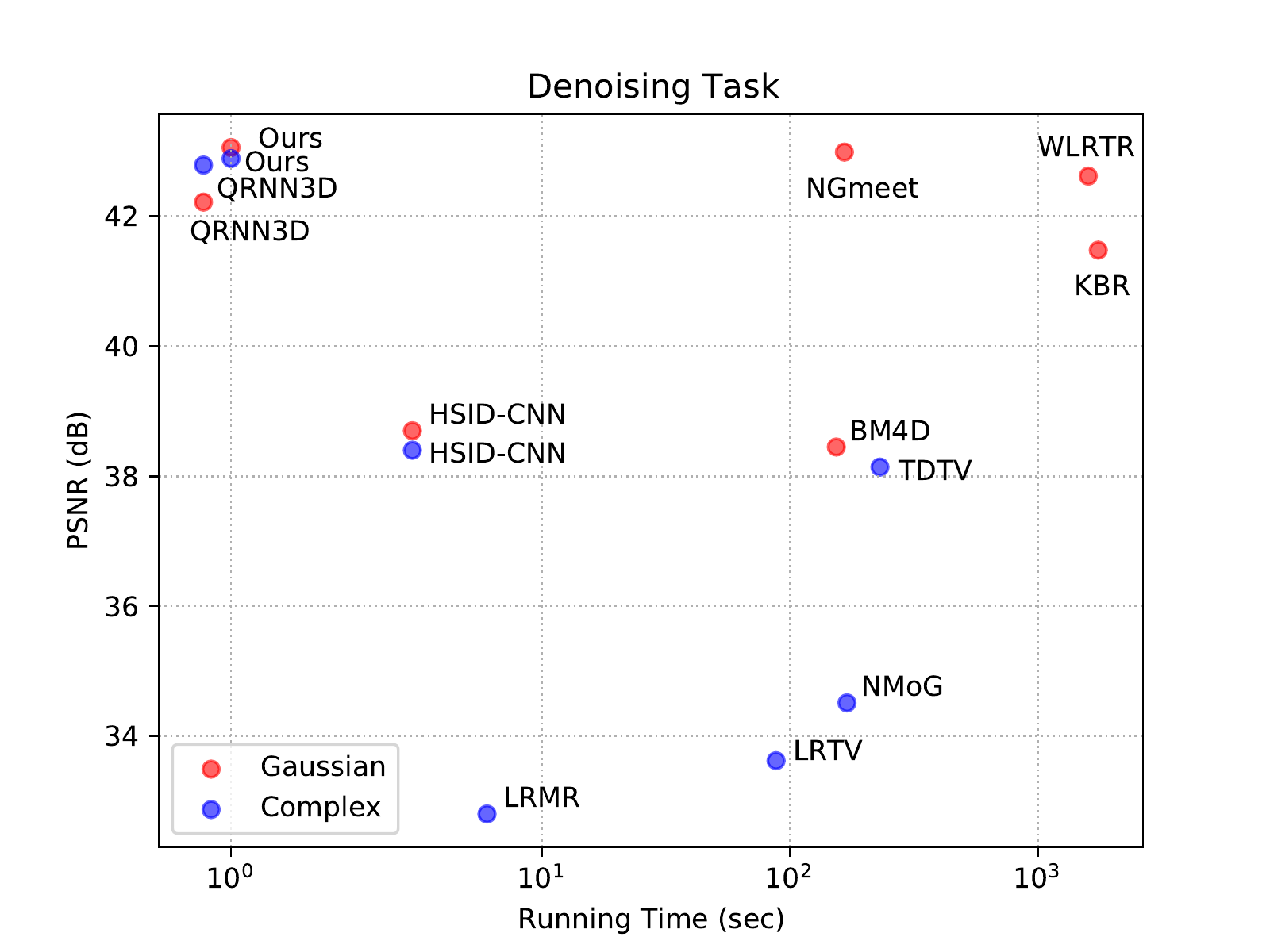}
    }
    \subfigure{
        \includegraphics[width=0.48\linewidth]{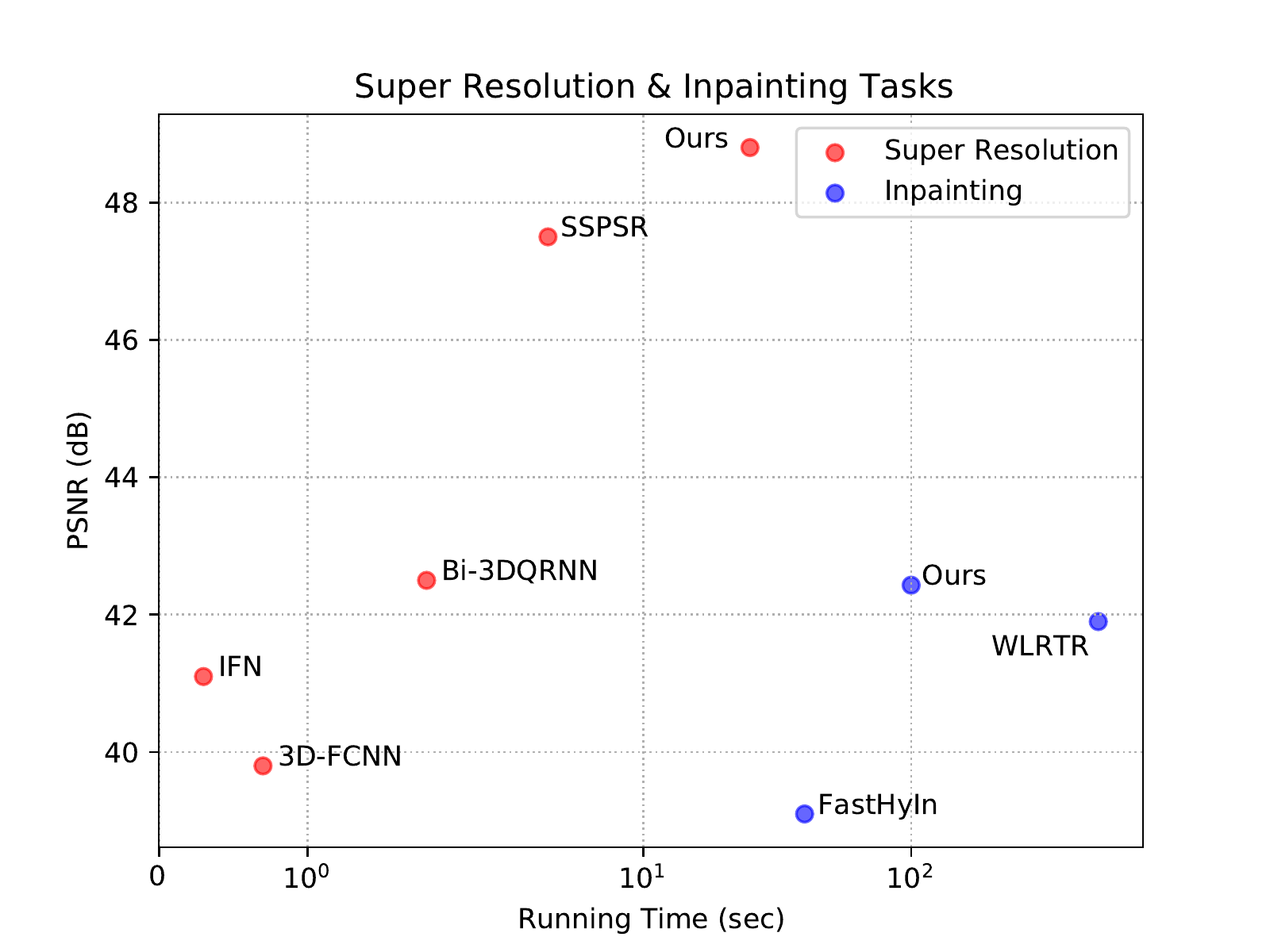}
    }
    \caption{The computational complexity of different competing methods for denoising, super-resolution, and inpainting tasks. Since the source code of NGmeet for compressed sensing is not available, the results of compressed sensing are not included.}
    \label{fig:time-com}
\end{figure*}

\subsubsection{Comparison with Traditional PnP Denoisers}

In order to prove the superiority of our denoiser within the PnP framework against traditional Gray/RGB PnP denoiser, we conduct an experiment on super-resolution for two widely used state-of-the-art denoisers in grayscale image restoration, \ie, IRCNN \cite{zhang2017beyond} and DRUnet \cite{zhang2020plug}. By denoising each band of HSIs sequentially, these denoisers are plugged into our plug-and-play framework for evaluation. As shown in Table \ref{tab:super-resolution-aba}, our denoiser is consistently better than the competing grayscale denoisers in terms of all metrics, indicating it serves as a better plug-and-play prior for HSIs. The visual comparison in Figure \ref{fig:cmp2d-result} also verifies the superiority of our method. It can be easily observed that the restored images by grayscale denoiser are blurred and introduce evident artifacts, while our method produces clearer results without evident artifacts.

\begin{table}[h!]
    \caption{Quantitative comparison against other PnP denoisers on super-resolution task and ICVL dataset. All the denoisers are directly plugged into the same PnP framework to evaluate the performance. }

    \begin{center}
        \small
        \begin{tabular}{@{}ccccc@{}}
            \toprule
            \multirow{2}{*}{Scale Factor} & \multicolumn{1}{c}{\multirow{2}{*}{Metric}} & \multicolumn{3}{c}{Denoiser}                                               \\ \cmidrule(l){3-5}
                                          & \multicolumn{1}{c}{}                        & IRCNN\cite{zhang2017beyond}  & DRUNet\cite{zhang2020plug} & Ours           \\ \midrule
            \multirow{3}{*}{$\times2$}    & PSNR                                        & 35.15                        & 36.09                      & \textbf{38.12} \\
                                          & SSIM                                        & 0.946                        & 0.953                      & \textbf{0.973} \\
                                          & SAM                                         & 0.124                        & 0.105                      & \textbf{0.059} \\ \bottomrule
            \multirow{3}{*}{$\times8$}    & PSNR                                        & 30.11                        & 30.61                      & \textbf{32.51} \\
                                          & SSIM                                        & 0.886                        & 0.901                      & \textbf{0.927} \\
                                          & SAM                                         & 0.154                        & 0.169                      & \textbf{0.077} \\ \bottomrule
        \end{tabular}
    \end{center}
    \label{tab:super-resolution-aba}
\end{table}

\subsubsection{Comparisons on More Datasets}

In order to test the generalizability of our plug-and-play method over different scenes, we conduct a series of experiments for HSI inpainting on other three HSI datasets, including two natural HSI datasets, \ie, CAVE \cite{CAVE_0293} and Harvard \cite{harvard}, and one remote-sensed dataset, \ie, Pavia University \cite{paviaU}. It should be noted that we directly apply the denoiser trained on the ICVL dataset for these experiments. As shown in Table \ref{tab:destripe-more-datasets}, our method outperforms the competing methods even though these scenes are completely unseen by the denoiser, suggesting the strong robustness of our method.

\begin{table}[h]
    \caption{Quantitative inpainting results (PSNR) of different methods on the CAVE, Harvard, and Pavia University datasets.}

    \begin{center}
        \small
        \begin{tabular}{@{}ccccc@{}}
            \toprule
            Method  & Noisy & WLRTR\cite{chang2020weighted} & FastHyIn\cite{zhuang2018fast} & Ours           \\ \midrule
            CAVE    & 14.22 & 35.36                         & 33.93                         & \textbf{37.00} \\
            Harvard & 14.20 & 38.27                         & 36.93                         & \textbf{39.37} \\
            PaviaU  & 14.22 & 32.22                         & 31.76                         & \textbf{32.73} \\ \midrule
        \end{tabular}
    \end{center}
    \label{tab:destripe-more-datasets}
\end{table}

\begin{table}[]
    \small
    \caption{Quantitative evaluation of the proposed PnP approach against competing deep-learning-based methods by the number of parameters (Params) and PSNR. The number of parameters under different settings is the same, and we provide the PSNR results on two settings for comparison, \ie, mixed noise levels of [30,70] for Gaussian denoising, and $4\times$ scale factor for super-resolution.}
    \label{tab:params}
    \begin{center}
        \setlength{\tabcolsep}{0.4cm}
        \begin{tabular}{@{}cccc@{}}
            \toprule
            Task                                        & Model                                & Params & PSNR  \\ \midrule
            \multirow{3}{*}{Denoising}                  & HSID-CNN\cite{yuan2018hyperspectral} & 0.40M  & 37.80 \\
                                                        & QRNN3D\cite{wei20203}                & 0.84M  & 41.37 \\
                                                        & Ours                                 & 14.28M & 42.23 \\ \midrule
            \multirow{4}{*}{\begin{tabular}[c]{@{}c@{}}Super \\ Resolution\end{tabular}} & IFN\cite{hu2020hyperspectral}        & 0.17M  & 38.92 \\
                                                        & SSPSR\cite{fu2021biqrnn3d}           & 20.82M & 39.20 \\
                                                        & Bi-3DQRNN\cite{fu2021biqrnn3d}       & 1.29M  & 39.56 \\
                                                        & Ours                                 & 14.28M & 40.96 \\ \bottomrule
        \end{tabular}
    \end{center}
\end{table}
\subsubsection{Computational Efficiency}

In this section, we analyze and compare the computational efficiency of our method and the competing classical and deep-learning-based methods. As indicated in Algorithm \ref{alg:pnp}, our method is iterative based, and most of the computation lies in the forward computation of deep denoiser since the data-fidelity term can generally be solved in closed-form. Hence, the approximate time complexity of our method is $O(nD)$ where $n$ is the number of iteration and $D$ is the inference time of denoiser. For denoising task, $n=1$. For super-resolution, compressed sensing, and inpainting, $n$ is set to 25, 50, 100, respectively.

Figure \ref{fig:time-com} shows the approximate running time of different methods on denoising, super-resolution, and inpainting tasks. All the tests are run on an i9-10850k CPU and an Nvidia RTX 3090 GPU. It can be seen that our method is significantly faster than most traditional methods while preserving superior performance for denoising tasks. QRNN3D is slightly faster than ours at the cost of slight performance degradation. For super-resolution, our method is training-free and achieves the best performance, but is slower than the competing pure deep-learning-based methods, which are all trained specifically. Table \ref{tab:params} also provides the quantitative comparison by the number of parameters on the denoising and super-resolution task. For inpainting tasks, all the methods are training-free. It can be observed that our method is significantly faster than WLRTR and achieve better performance. FastHyIn is faster than ours, but its performance also degrades severely. Overall, when compared with traditional methods, our method is significantly faster than most of them and achieves better performance. Simultaneously, when compared with specifically trained deep-learning-based methods, our method can also achieve better performance without any specific training at the cost of a tiny increase in running time.

\section{Conclusion}\label{sec:conclude}

In this paper, we develop a flexible and effective PnP-ADMM approach for HSI restoration, in which a novel deep HSI denoiser is introduced as an implicit PnP image prior. With careful design, our denoiser not only achieves the state-of-the-art performance in HSI denoising task by properly exploiting the intrinsic characteristics underlying HSIs, but also, acting as a powerful HSI prior, finely supports the plug-and-play framework to solve different HSI restoration tasks without any additional training. The extensive experimental results and analysis show that our method is able to achieve superior generalizability and competitive or even better performance against the learning-based methods in a variety of HSI restoration problems.

\section*{Acknowledgement}

This work was supported by the National Natural Science Foundation of China under Grants No. 62171038, No. 61827901, and No. 62088101.

\bibliographystyle{elsarticle-num}
\bibliography{main}

\end{document}